\documentclass[a4paper,12pt]{article}

\usepackage{jheppub} 
\usepackage{slashed}
\usepackage{mathtools}
\usepackage{amsfonts}
\usepackage{subcaption}
\usepackage{comment}
\usepackage{multirow}
\usepackage{array}
\usepackage{hhline}
\usepackage{booktabs}
\usepackage[T1]{fontenc} 

\pdfoutput=1 

\title{\boldmath Dielectric constant of holographic insulator 
	from quantum geometry}
\author[1]{Debabrata Ghorai,}
\author[2]{Taewon Yuk,}
\author[2]{Sang-Jin Sin}
\affiliation[1]{Department of Education, IIT Kharagpur, West Bengal, India}
\affiliation[2]{Physics Department, Hanyang University, Seoul, Korea}

\emailAdd{debabrataghorai.phy@gmail.com}
\emailAdd{tae1yuk@gmail.com}
\emailAdd{sangjin.sin@gmail.com}

\abstract{We study dielectric  properties of holographic insulators  from the quantum geometry of the ground state. We  realize  the   insulating state  by the pseudoscalar interaction of holographic fermions  and calculate  the fermionic Green function  to  
	extract the effective   hamiltonian. We then derive the quantum metric and Berry curvature to  compute the transports and finally get the dielectric constant  from the imaginary part of conductivity. 
}

\begin{document} 
	\maketitle
	\flushbottom
	
	\section{Introduction}
	The geometric structure of quantum states has emerged as a central concept in modern condensed matter physics, driven largely by the geometric and topological phenomena observed in quantum materials  \cite{QGTRachel, QGTAhn, Wang1, RestaR}. The \textit{Quantum Geometric Tensor} (QGT) 
	encodes both the \textit{quantum metric  }  and   the \textit{Berry curvature}  as its real and imaginary parts respectively \cite{Provost:1980nc,10.1098/rspa.1984.0023}. These quantities together provide a powerful geometric framework to describe the response of a quantum system under  external sources as well as the intrinsic topological quantities.   
	
	The central question here  is how to encode the interaction effects for quantum materials while 
	quantum geometry  as well as the linear response theory 
  is formulated using 
 the band structure  which does not capture the effects of   interaction. 
 In this paper we try to use the holographic approach to overcome this problem. 
		
	The gauge/gravity duality \cite{MaldacenaM, GKPW1, GKPW2},  has been used to describe phenomena in strongly coupled systems \cite{HQMReV}. In ref. \cite{Iqbal:2009fd}, fermionic spectral functions for different types of interactions have been studied both analytically and numerically over the last decade. We have investigated  various  systems including  holographic Mott insulators \cite{Ghorai:2024nxs}, Fermi arc in holographic superconductors \cite{Ghorai:2023vuo, Ghorai:2023wpu}, Kondo physics \cite{Han:2024rbr}.  Here 
	holographic duality  incorporates electron-electron   interactions via AdS bulk geometry and capture the strong coupling effects \cite{SangJinNature}. Therefore, we expect that  if quantum geometric formalism can be formulated  in holography, it would yield a fruitful results for strongly coupled system.  However, there is still no complete method  to compute  transport coefficients utilizing the holographic fermion spectral functions directly: 
 although the holographic bulk fermion one-loop  calculation had been done in \cite{Faulkner:2013ica}, the method is   difficult to  use or verify  due to the convergence issues in the self-energy and vertex corrections.  
 
 To overcome those problem, we   propose a  new  method to compute transport directly from the fermion Green's function 
 by extending the band structure: 	
	instead of   single-particle band, we model the system's internal degrees of freedom---such as sublattice pseudospins or energy branches---using bulk fermionic multiplets that naturally encode both inter-component and many-body self-interactions through the dual gravitational geometry. 
	Such degrees of freedom make sense in the holographic fermion as well as the non-interacting fermions.  The nature of  degrees of freedom can be  understood   in the limit of $m\to 1/2$ where it is known that the bulk fermion's behavior follows that of Fermi liquid \cite{Oh:2021xbe}.  
	Given the   retarded Green's function of   coupled two flavored fermions, we can construct an effective Hamiltonian, 
	   from which we can construct the QGT.  To realize an  insulator, we consider the bulk fermion which is known to have gap\cite{Oh:2021xbe}.   
We will compute   conductivity  using the modified Kubo formula for holography, which works  for gapped case. The result exhibits phase transition as   frequency increases at zero temperature,   signaling  the presence of a quantum phase transition. 
		For finite temperature, it shows 
		insulator to metal phase transition in ac conductivity.
	 
	In summary, we propose a novel method to compute  transport properties alongside fermionic spectral functions. Our approach allows   calculation of holographic transport  without solving the bulk gauge field equations,  providing  a unique platform for evaluating  all the quantities 
	such as capacitance,  dielectric constants,  transport, topology and quantum geometry for 
 the  holographic insulator   starting from fermion.

	This paper is organized as follows. In Sec.~2, we introduce the basic setup and outline the method for computing the fermionic spectral function and the effective Hamiltonian in the holographic framework. In Sec.~3, we compute the QGT for the holographic insulating system. In Sec.~4, we analytically study the transport properties at zero temperature. In Sec.~5, we compare the finite-temperature holographic results with the corresponding field-theory results. 
	Finally, we summarize our main results, compare them with experimental observations, and discuss possible future directions.

	\section{Holographic spectral function and effective Hamiltonian}
	The fermionic action for insulating (gapped) system \cite{Oh:2020cym} reads 
	\begin{eqnarray}
		S_{\psi} = \int d^4x \sqrt{-g} \left[i \bar{\psi} (\Gamma^{\mu}D_{\mu}- m_{f} - i \eta \phi(z) \Gamma^{5}) \psi \right]  + S_{bdy}, \\ ~\text{ where } ~ S_{bdy} = i \int d^3x  \sqrt{-h} \bar{\psi}\psi ~.
	\end{eqnarray}
	Here, $\eta$ represents the coupling constant and $\phi(z)=M_5 z$ contains only external source representing some  lattice effect creating a gap. 
	The spinor's covariant derivative is denoted by $D_{\mu}=\partial_{\mu} + \frac{1}{4} \omega_{\mu\bar{\alpha\beta}}\Gamma^{\bar{\alpha\beta}}$. The bulk gamma matrices  used in  this paper are  as follows:
	\begin{eqnarray}
		\Gamma^{\underline{t}} = -i \sigma_1 \otimes \sigma_0 \quad
		\Gamma^{\underline{x}} = \sigma_2 \otimes \sigma_1, \quad
		\Gamma^{\underline{y}} = \sigma_2 \otimes \sigma_2, \quad
		\Gamma^{\underline{z}} = \sigma_3 \otimes \sigma_0, \quad 
		\Gamma^{5} = \sigma_2 \otimes \sigma_3, ~~~~
	\end{eqnarray} 
	where underlined indices represent tangent space ones.
	The Dirac equation in the bulk is 
	\begin{eqnarray}
		(\Gamma^{\mu}D_{\mu}- m_{f}-i\eta M_5 \Gamma^{5} z)\psi = 0   ~.
	\end{eqnarray} 
	Introducing   $\Phi$ by 
	\begin{eqnarray}
		\psi (t, x, y, z) = \frac{1}{(-gg^{zz})^{1/4}} e^{-i \omega t + i k_x x + i k_y y } \Psi(z) ~~.
		\label{fanstz}
	\end{eqnarray}
	it allows us to eliminate the spin connection term and eq. (\ref{fanstz})  is simplified to give an explicit form:
	\begin{align}
		&\ \left[\Gamma^{\underline{z}}\partial_z - i  \sqrt{\frac{g^{tt}}{g^{zz}}}\omega \Gamma^{\underline{t}}-\frac{m_f}{\sqrt{g^{zz}}}\right]\Psi(z)  \nonumber\\
		&\ + i \left(
		\sqrt{\frac{g^{xx}}{g^{zz}}} k_x \Gamma^{\underline{x}} +  \sqrt{\frac{g^{yy}}{g^{zz}}} k_y \Gamma^{\underline{y}}\right) \Psi(z) - \frac{i\eta M_5 z}{\sqrt{g^{zz}}}\Gamma^{5} \Psi(z) = 0.
		\label{FDiracEq}
	\end{align} 
	In this section, we will calculate the density of states by solving the Dirac equation.  For this, we express the four-component spinor as $\Psi(z)=\left(\Psi_{+}(z), \Psi_{-}(z)\right)^T $ where $\Psi_{\pm}= \left(\Psi_{\pm 1}, \Psi_{\pm 2}\right)$. 
	Then, the Dirac equation becomes 
	\begin{eqnarray}
		\left[\partial_z \mp \frac{m_f}{\sqrt{g^{zz}}}\right] \Psi_{\pm}(z) =\pm  \left[  K_{j}\gamma^{j}\Psi_{\mp}(z) \right] \pm \frac{\eta M_5 z}{\sqrt{g^{zz}}} \sigma_3 \Psi_{\mp}(z)   
		\label{eq39}
	\end{eqnarray}
	\begin{eqnarray}
		\text{ where }  K_{j}= \left(\sqrt{\frac{g^{tt}}{g^{zz}}} \omega, -\sqrt{\frac{g^{xx}}{g^{zz}}}k_x, -\sqrt{\frac{g^{yy}}{g^{zz}}}k_y\right), \; \gamma^{j}=\left(\sigma_0, \sigma_1, \sigma_2\right).
	\end{eqnarray}
	In the asymptotic    region  $z\rightarrow 0$,  $g^{\mu\nu}\rightarrow z^2 \eta^{\mu\nu}$ with   the Minkowski metric $\eta^{\mu\nu}$. In the asymptotic limit, the source and condensation are given for $|m_f|<\frac{1}{2}$ by \cite{Ghorai:2023wpu}
	\begin{eqnarray}
		\Psi_{+}(z)  \overset{z\rightarrow 0}{\simeq}\mathbf{A} z^{m_f}, \quad
		\Psi_{-}(z)  \overset{z\rightarrow 0}{\simeq}\mathbf{D} z^{-{m_f}} ~~~.
		\label{eqnu18}
	\end{eqnarray}
	
	Following the   procedure in our earlier work \cite{Ghorai:2023vuo}, we can write down the boundary action in the following form
	\begin{eqnarray}
		S_{bdy} = \int d^3x  \left[ \Psi_{-}^{\dagger}(z) \tilde{\Gamma} \Psi_{+}(z)+ \Psi_{+}^{\dagger}(z) \tilde{\Gamma}\Psi_{-}(z) \right]_{z\to 0}
		\label{eq218}
	\end{eqnarray}
	where the boundary gamma matrix $\tilde{\Gamma}= \sigma_0$. Recasting the Dirac equation, the flow equation for bulk Green's function $\mathbb{G} (z)$  is derived in Appendix A.1.  Using the horizon value of the $\mathbb{G} (z)$ as `initial condition' we can solve the flow equation, to numerically calculate the bulk  Green's function $\mathbb{G} (z)$ near the boundary from which the retarded Green's function $\mathbb{G}_{R}$ can be obtained using the   relation
	\begin{eqnarray}
		\mathbb{G}_{R} (k_x,k_y,\omega) = \lim_{z\to 0} U(z) \mathbb{G}(z) U(z) 
	\end{eqnarray}
	where $U(z)= \text{diag}(z^{m_f}, z^{m_{f}})$.
	The spectral function (SF) is defined as
	\begin{eqnarray}
		A (k_x, k_y, \omega) = \text{Tr}(\text{Im}(\mathbb{G}_{R}(k_x,k_y,\omega))) ~.
	\end{eqnarray}
	This fermionic spectral function gives the generalized concept of the electronic structure which contains information of the fuzziness of the dispersion relation as well as how degrees of freedom moved from the non-interacting case. Notice that  these are information of the boundary theory although we calculated them from the bulk quantities. 
	
	Knowing the  retarded Green's function $\mathbb{G}_{R}$, we can compute the so called 'topological Hamiltonian'  which determines the  state information \cite{Wang1}.  It is defined by \footnote{This comes from $ G(\omega)  = \frac{1}{\omega - H_{I}} \quad \Rightarrow H_{I} = \omega - G^{-1}(\omega) \Rightarrow
		H_{T} =  - G^{-1}(\omega=0)$, where $H_{I}$ is bulk (material) interacting Hamiltonian.} 
	\begin{eqnarray}
		H_{T}(k_x,k_y) = - \mathbb{G}^{-1}_{R}(k_x,k_y,0)~.
	\end{eqnarray}
	We can now diagonalize   above Hamiltonian to   get the    states of the system, from which  we can compute quantum metric and Berry curvature. Since we are working in one flavor fermion in holographic setup, the boundary retarded Green's function is two by two matrix which leads to a topological Hamiltonian with rank 2. We can write this topological Hamiltonian in terms of Pauli basis as following form:
	\begin{align}
		H_T(k_x,k_y) = h_0(\mathbf{k}) \, \mathbb{I}_2 + \mathbf{h}(\mathbf{k}) \cdot \boldsymbol{\sigma},
	\end{align}
	where  $\mathbf{k} = (k_x, k_y)$, 
	$\mathbf{h} = \big(h_1(\mathbf{k}), h_2(\mathbf{k}), h_3(\mathbf{k})\big)$, 
	$\boldsymbol{\sigma} = (\sigma_1, \sigma_2, \sigma_3)$, and the $\sigma_i$'s are the Pauli matrices.
	The quantum metric and Berry curvature for ground state read (see the Appendix \ref{a2}.)
	\begin{align}
		g_{\mu\nu} = \frac{1}{4}\partial_{\mu}\hat{\mathbf{h}}.\partial_{\nu}\hat{\mathbf{h}}, \quad F_{\mu\nu}= \frac{1}{2} \hat{\mathbf{h}}.\partial_{\mu}\hat{\mathbf{h}}\times \partial_{\nu}\hat{\mathbf{h}}=\frac{1}{2\hat{h}_3}\left(\partial_{\mu}\hat{h}_1 \partial_{\nu}\hat{h}_2 - \partial_{\nu}\hat{h}_1 \partial_{\mu}\hat{h}_2 \right) 
	\end{align}
	where the normal vector $\hat{\mathbf{h}} = \frac{\mathbf{h}}{|\mathbf{h}|}$.

	\section{Quantum geometric tensor for holographic   system}
\begin{figure}[h!]
		\centering
		\includegraphics[scale=0.35]{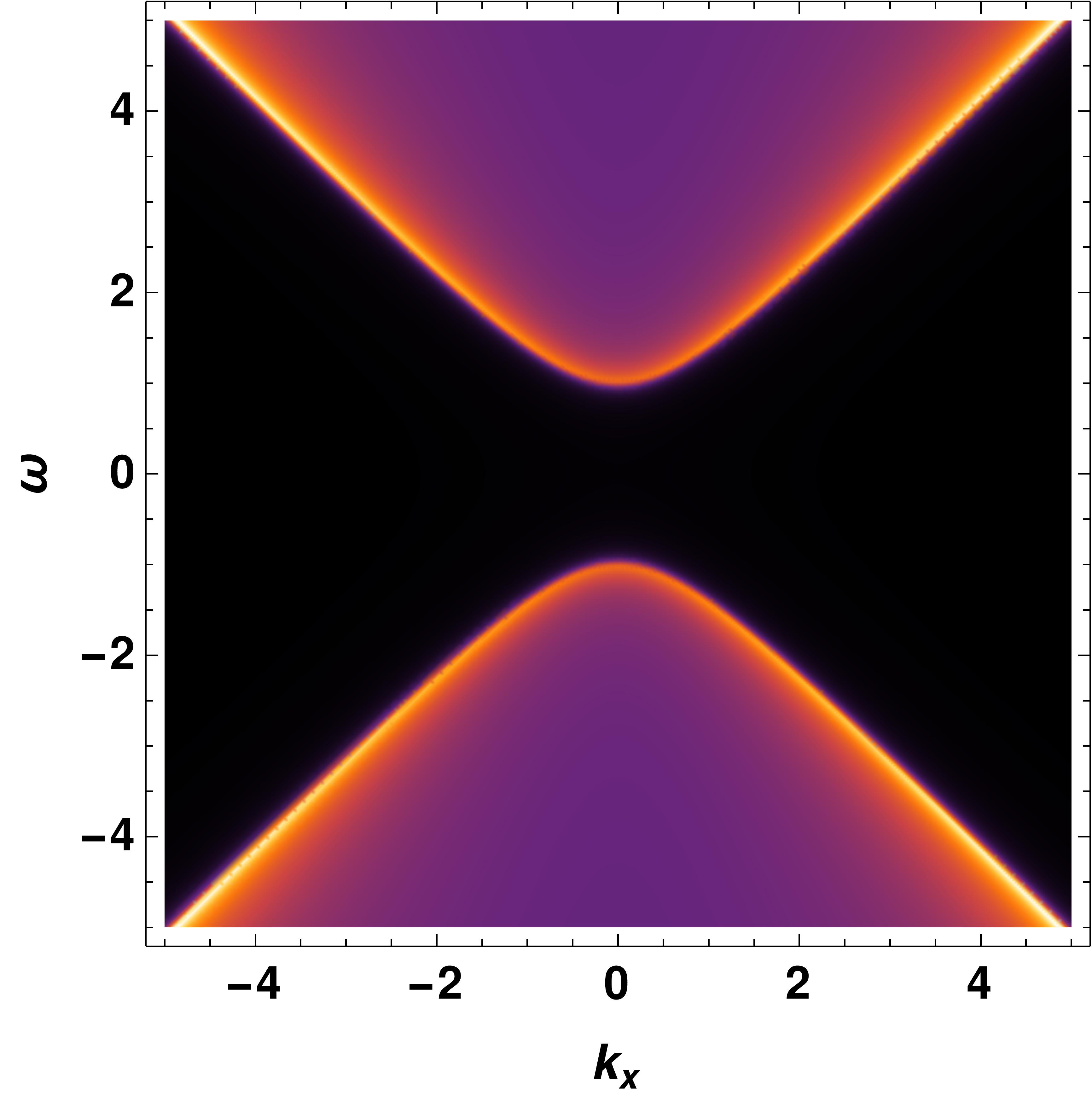}
		\caption{Spectral function at $T=0$ with $M_5=1$}
		\label{sfun}
	\end{figure}

	To determine the QGT for the given system, we first compute the retarded Green's function by solving Eq.~(\ref{eq39}) using the method developed in \cite{Oh:2020cym, Sukrakarn:2023ncp} for a probe background geometry. At finite temperature, we primarily consider the AdS-Schwarzschild black hole background, whereas for the zero-temperature we employ the pure AdS geometry. Such approximation allows analytic power to trace what is going on in each step although we would not trust the details of dynamical aspects in low energy.   For the pure  AdS limit, the retarded Green's function can be obtained analytically 
	In  this case,  for
	if we set $m_f=0$  \cite{Sukrakarn:2023ncp}: 
	\begin{eqnarray}
		\mathbb{G}_{R} (k_x,k_y,\omega) = \left(
		\begin{array}{cc}
			\frac{\omega -M_5}{\sqrt{k_x^2+k_y^2+M_5^2-\omega ^2}} & \frac{k_x-i k_y}{\sqrt{k_x^2+k_y^2+M_5^2-\omega ^2}} \\
			\frac{k_x+i k_y}{\sqrt{k_x^2+k_y^2+M_5^2-\omega ^2}} & \frac{M_5+\omega }{\sqrt{k_x^2+k_y^2+M_5^2-\omega ^2}} \\
		\end{array}
		\right),
	\end{eqnarray}
	which results in the fermionic spectral function  $ A (k_x, k_y, \omega)= Im\left(\frac{2\omega}{\sqrt{k_x^2+k_y^2+M_5^2-\omega ^2}}\right)$. This will gives us the dispersion relation and band structure in the holographic setup, which is shown in Figure \ref{sfun}. 
	We can now calculate its  topological Hamiltonian to get 
	\begin{eqnarray}
		H_T(k_x,k_y) =  \left(
		\begin{array}{cc}
			\frac{M_5}{\sqrt{k_x^2+k_y^2+M_5^2}} & \frac{-k_x+i k_y}{\sqrt{k_x^2+k_y^2+M_5^2}} \\
			\frac{-k_x-i k_y}{\sqrt{k_x^2+k_y^2+M_5^2}} & \frac{-M_5 }{\sqrt{k_x^2+k_y^2+M_5^2}} \\
		\end{array}
		\right) ~.
	\end{eqnarray} 
	From this, we can obtain the   states which yields the quantum geometry and the Berry curvature as follow:
	\begin{eqnarray}
		g = \left(
		\begin{array}{cc}
			\frac{k_y^2+ M_5^2}{4(k^2+M_5^2)^2} & \frac{-k_x k_y}{4(k^2+M_5^2)^2} \\
			\frac{-k_x k_y}{4(k^2+M_5^2)^2} & \frac{k_x^2+ M_5^2}{4(k^2+M_5^2)^2} \\
		\end{array}
		\right), \quad F = \left(
		\begin{array}{cc}
			0 & \frac{M_5}{2(k^2+M_5^2)^{\frac{3}{2}}} \\
			-\frac{M_5}{2(k^2+M_5^2)^{\frac{3}{2}}} & 0 \\
		\end{array}
		\right)
	\end{eqnarray}
	where $k=\sqrt{k_x^2+k_y^2}$. From this Berry curvature, we can define topologically invariant quantity, the Chern number  as 
	\begin{eqnarray}
		\mathcal{C} = \frac{1}{(2\pi)^2}\int F_{xy} dk_x dk_y = \frac{1}{(2\pi)^2}\int^{\infty}_{0} k dk\int_{0}^{2\pi} F_{xy}  d\theta = \frac{{\rm sign}(M_5)}{4\pi} ~. 
	\end{eqnarray}
	The trace of the quantum metric is
	$$\frac{k^2+2M_5^2}{4(k^2+M_5^2)^2},$$ which will be used to compute the transport properties. This is   zero temperature result. 
	The QGT is a gauge-invariant quantity, and the holographic QGT is found to be identical to that of a two-dimensional Dirac Hamiltonian upon the identification $m=M_5$. Conversely, the exact agreement of these gauge-invariant quantities suggests that the holographic parameter $M_5$ can be interpreted as the Dirac mass $m$. 
	
	The fact that QGT depends only on the ratio $\hat h$ rather than $h$ washes out much of the interaction effect and this is responsible for why the topology of the interacting system is the same as that of the non-interacting one.  The stability of the topology was already observed based on the full numerical calculation in \cite{Song:2019asj}. What 
	we get here  is understanding the analytic reason of the stability.  
	However, 	this also  reveals the character of  the metric tensor  on revealing the   dynamics of the system:  although it is not completely topological, it is semi-topological  because 
	both $g_{\mu\nu}$ and $F_{\mu\nu}$ depends only on $\hat h$. 
	Nevertheless, it is still interesting to see how much dynamics is contained in the  metric    which is not completely topological.       Whether the semi-topological nature of the quantum metric would continue to hold to more than two bands is an open question.  
	
	
	\section{Transport   at zero temperature} 
	In the band theory framework, 
	the Kubo formula for interband conductivity can be expressed as follows \cite{QGTRachel}:
	\begin{eqnarray}
		\sigma_{\mu\nu}(\omega) &=& -\frac{2 i e^2}{\hbar}\sum_{ n\neq m}\int_{\text{BZ}}  \frac{\tilde{f}_{nm} \omega_{nm}\omega}{\omega_{nm}^2-\omega^2} g_{\mu\nu}^{nm} -\frac{e^2}{\hbar}\sum_{n\neq m}\int_{\text{BZ}}  \frac{\tilde{f}_{nm} \omega_{nm}^2}{\omega_{nm}^2-\omega^2} F_{\mu\nu}^{nm}
		\label{condequ1}
	\end{eqnarray}
	where $\tilde{f}_{nm}=f_n(1-f_m)$ and $f_n$ is Fermi-Dirac function for $E_n$ and $$\omega_{nm}(k)=\frac{E_n}{\hbar}-\frac{E_m}{\hbar} = \omega_n-\omega_m$$ is the energy difference between two bands. The details of derivation of Kubo formula in terms of geometry and topology is given in Appendix \ref{a3}. 
	
	For two band systems,   the interband QGT  for the  ground state $\psi_0$ and the excited state $\psi_1$ is given as follows: 
	\begin{eqnarray}
		Q^{00}_{\mu\nu} &=& \langle\partial_{\mu}\psi_0|\partial_{\nu}\psi_0\rangle -A^{00}_{\mu}A^{00}_{\nu} = \langle\partial_{\mu}\psi_0|\psi_1\rangle\langle\psi_1|\partial_{\nu}\psi_0\rangle = A^{01}_{\mu} A^{10}_{\nu} = Q^{01}_{\mu\nu} ~.
	\end{eqnarray}
	Notice that the matrix $A_\mu$ is hermitian. 
	When one of the  two   energy eigenvalues is negative,  the $\tilde{f}_{01}$ becomes $1$ in zero temperature limit, which simplifies the Kubo formula.  The conductivity matrix in this limit  becomes
	\begin{eqnarray}
		\sigma_{\mu\nu}(\omega) &=& -2 i \int_{\text{BZ}}  \frac{ \omega_{01}\omega}{\omega_{01}^2-\omega^2} g_{\mu\nu}^{01} -\int_{\text{BZ}}  \frac{ \omega_{01}^2}{\omega_{01}^2-\omega^2} F_{\mu\nu}^{01} ~,
		\label{eq43l}
	\end{eqnarray}
	with  $\frac{e^2}{\hbar}=1$. 
	In the low frequency limit one may write
	\begin{eqnarray}
		\sigma_{\mu\nu}(\omega) &=& 2 i\omega \int_{\text{BZ}}  \frac{ g_{\mu\nu}^{01}}{\omega_{10}}  -\int_{\text{BZ}}  F_{\mu\nu}^{01} = i\omega  \mathcal{G} \delta_{\mu\nu}- \mathcal{C} \epsilon_{\mu\nu}
	\end{eqnarray}
	where 
	\begin{eqnarray}
		\mathcal{G} = \int_{\text{BZ}} \frac{1}{\omega_{10}} Tr(g ^{01})= \int_{\text{BZ}} \frac{1}{\omega_{10}} Tr(g ^{00}) , 
			\end{eqnarray}
			 and 
			 	\begin{eqnarray}
			 \mathcal{C} = \int_{\text{BZ}} F^{01}_{xy}= \int_{\text{BZ}} F^{00}_{xy}~.
	\end{eqnarray}

	\subsection{Field theory results at zero temperature}
	
\begin{figure}[h!]
		\centering
		\begin{subfigure}[b]{0.45\textwidth}
			\centering
			\includegraphics[scale=0.3]{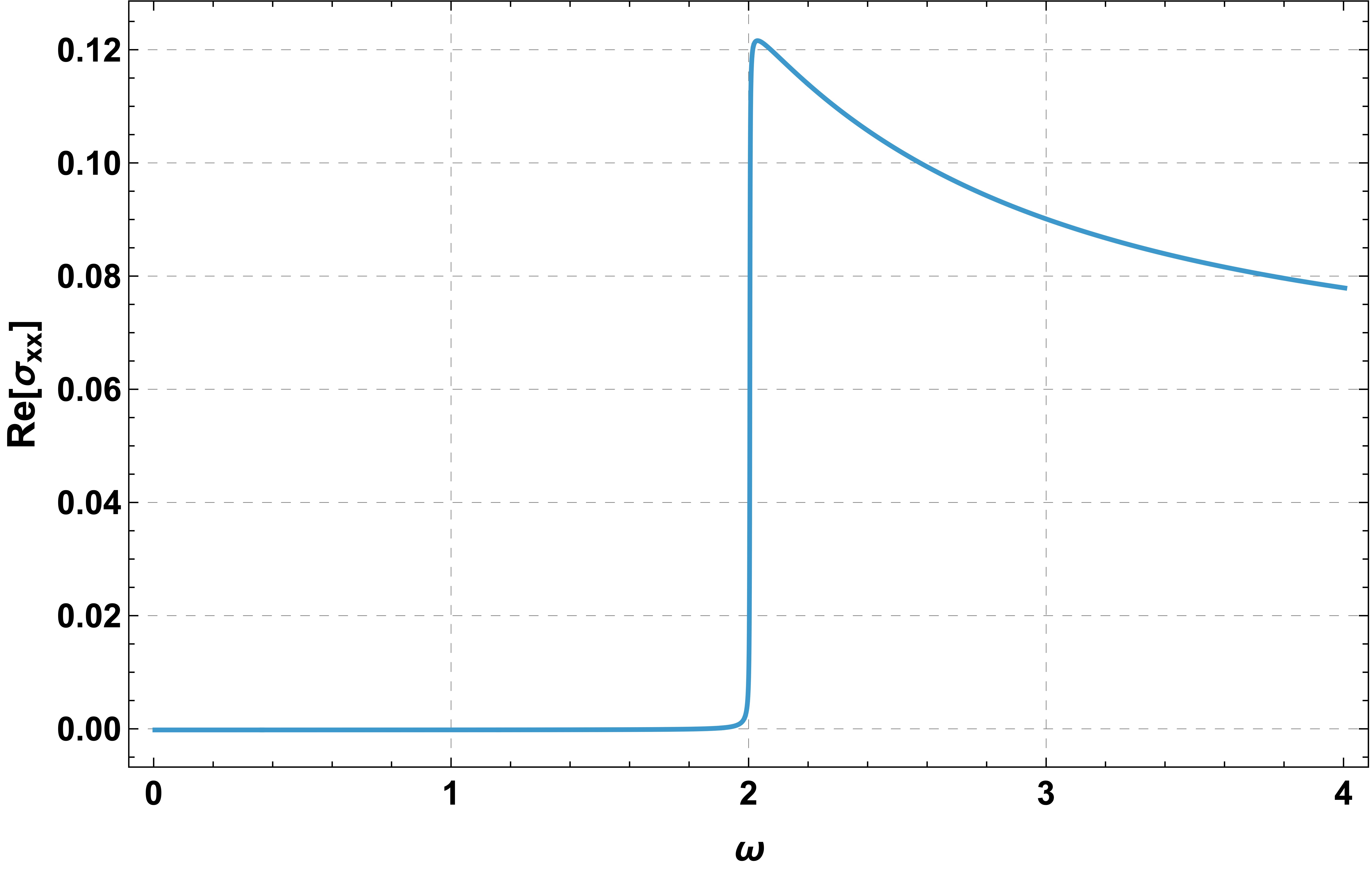}
			\caption{longitudinal conductivity $Re[\sigma_{xx}]$ }
		\end{subfigure}
		\hfil
		\begin{subfigure}[b]{0.45\textwidth}
			\centering
			\includegraphics[scale=0.3]{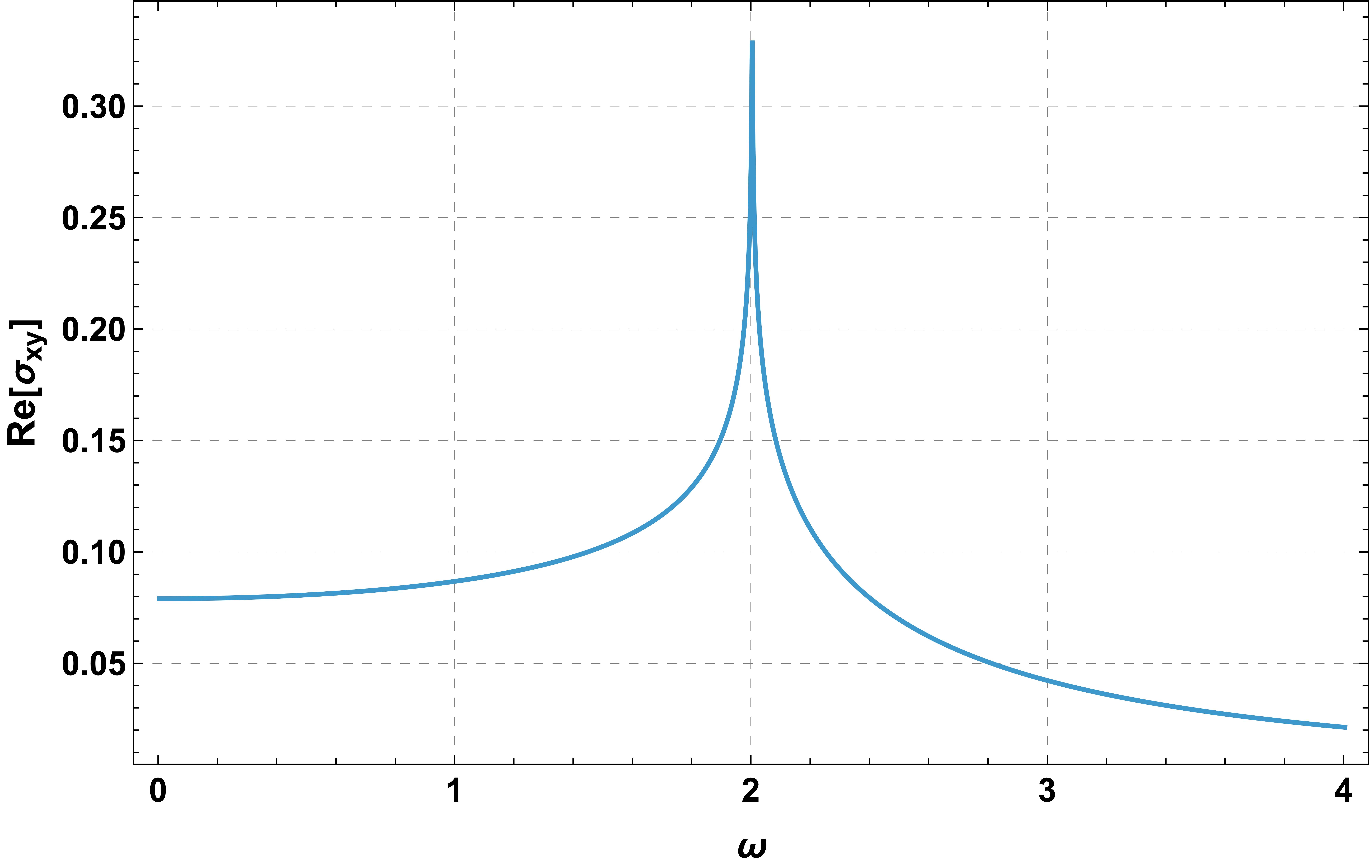}
			\caption{Hall conductivity $Re[\sigma_{xy}]$ }
		\end{subfigure}
		\caption{Transport for 2D Dirac Hamiltonian with gapping parameter $m=1$.}
		\label{FTCond0}
	\end{figure}

	In this section, we reproduce the zero-temperature transport properties following Ref.~\cite{QGTRachel}. The 2D Dirac Hamiltonian reads
	\begin{equation}
		H = k_x\sigma^x + k_y\sigma^y + m \sigma^z,
	\end{equation}
	which exhibits the same structure as the holographic model with a pseudoscalar interaction. Since this is a two-band system, we employ the general formalism (detailed in Appendix~\ref{Ase1}) to calculate the quantum geometric tensor (QGT), which yields
	\begin{equation}
		Q = \frac{1}{4(k^2+m^2)^2}\left(
		\begin{array}{cc}
			k_y^2+ m^2 & -k_x k_y +i m \sqrt{k^2+m^2}\\
			-k_x k_y -i m \sqrt{k^2+m^2} & k_x^2+ m^2 \\
		\end{array}
		\right).
	\end{equation}
	This result is identical to the holographic pseudoscalar interaction case under the identification $m = M_5$. From the energy dispersion relation, the energy difference between the upper and lower bands is $\omega_{10} = 2\sqrt{k^2 + m^2}$. At low frequencies, the conductivity is given by
	\begin{equation}
		\sigma_{\mu\nu}(\omega) = \frac{i \omega}{12\pi|m|} \delta_{\mu\nu} - \frac{\text{sgn}(m)}{4\pi}\epsilon_{\mu\nu},
	\end{equation}
	which is consistent with the previous findings in Ref.~\cite{QGTRachel}. Using Eq.~(\ref{eq43l}), we analytically compute the conductivity for the entire frequency range:
	\begin{align} 
		\sigma_{xx}(\omega) &= \frac{i}{4\pi}\left[ \left(\frac{1}{4}+\frac{m^2}{\omega^2}\right)\log\left(\frac{2|m|+\omega}{2|m|-\omega}\right)-\frac{|m|}{\omega}\right], \nonumber \\
		\sigma_{xy}(\omega) &= \frac{i m}{2\pi \omega} \cos^{-1}\left(\frac{2|m|}{\sqrt{4m^2-\omega^2}}\right).
	\end{align}
	We have shown the conductivity plots for 2D Dirac Hamiltonian with gapping parameter $m=1$ in the Figure \ref{FTCond0}. 
	Our field-theory expressions are fully consistent with the physical picture of an insulator-metal phase transition: when the incident frequency exceeds the energy-gap size, electrons can be excited from the valence band to the conduction band, resulting in a finite conductivity characteristic of the metallic state. In the next section, we discuss the zero-temperature holographic conductivity.

	\subsection{Holography results at zero temperature}
	
\begin{figure}[h!]
		\centering
		\begin{subfigure}[b]{0.3\textwidth}
			\centering
			\includegraphics[scale=0.25]{QGT/PScalarSF.jpeg}
			\caption{For holo int. $M_5=1$ }
		\end{subfigure}
		\hfil
		\begin{subfigure}[b]{0.3\textwidth}
			\centering
			\includegraphics[scale=0.35]{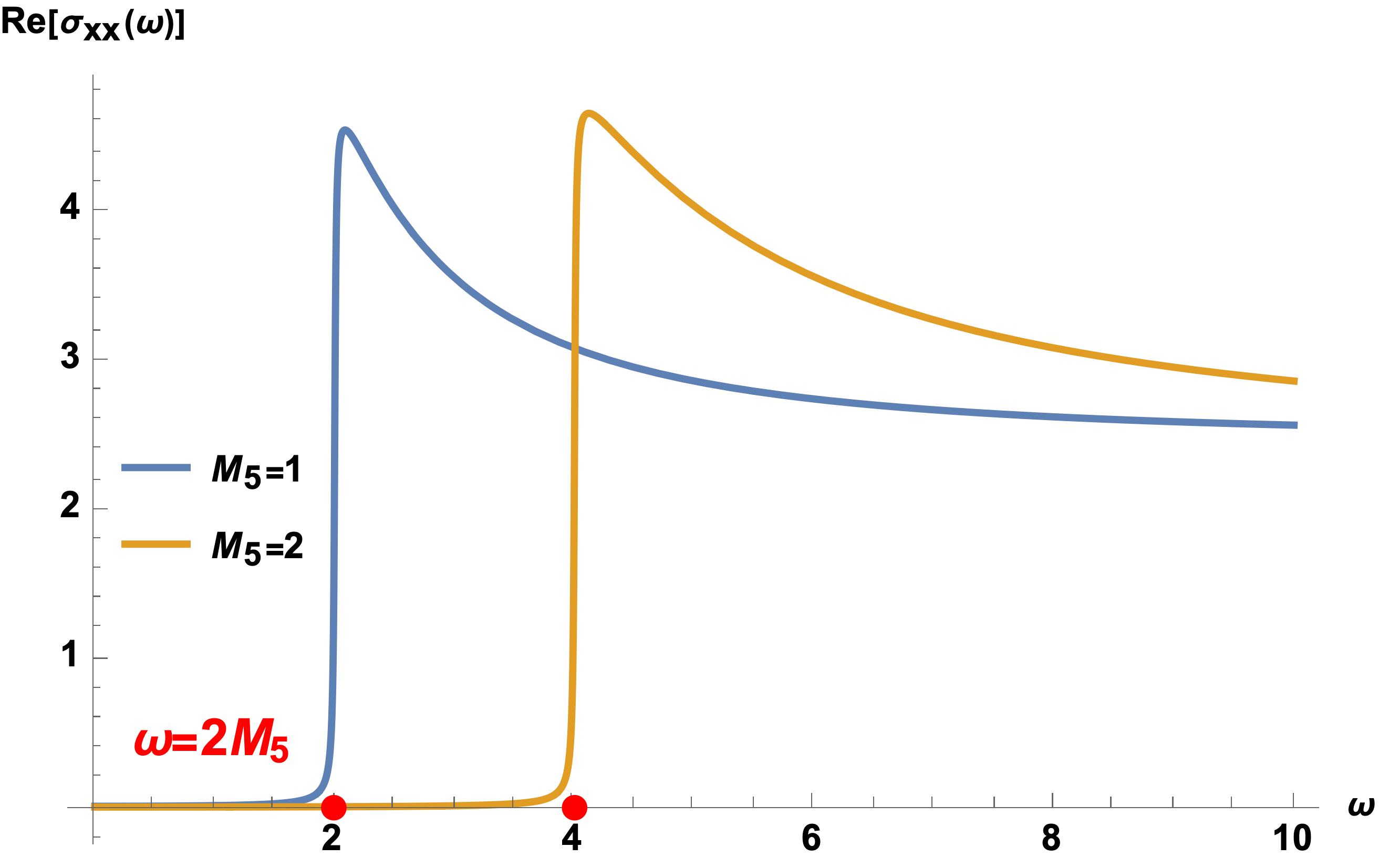}
			\caption{$Re[\sigma_{xx}(\omega)]$}
		\end{subfigure}
		\hfil
		\begin{subfigure}[b]{0.3\textwidth}
			\centering
			\includegraphics[scale=0.35]{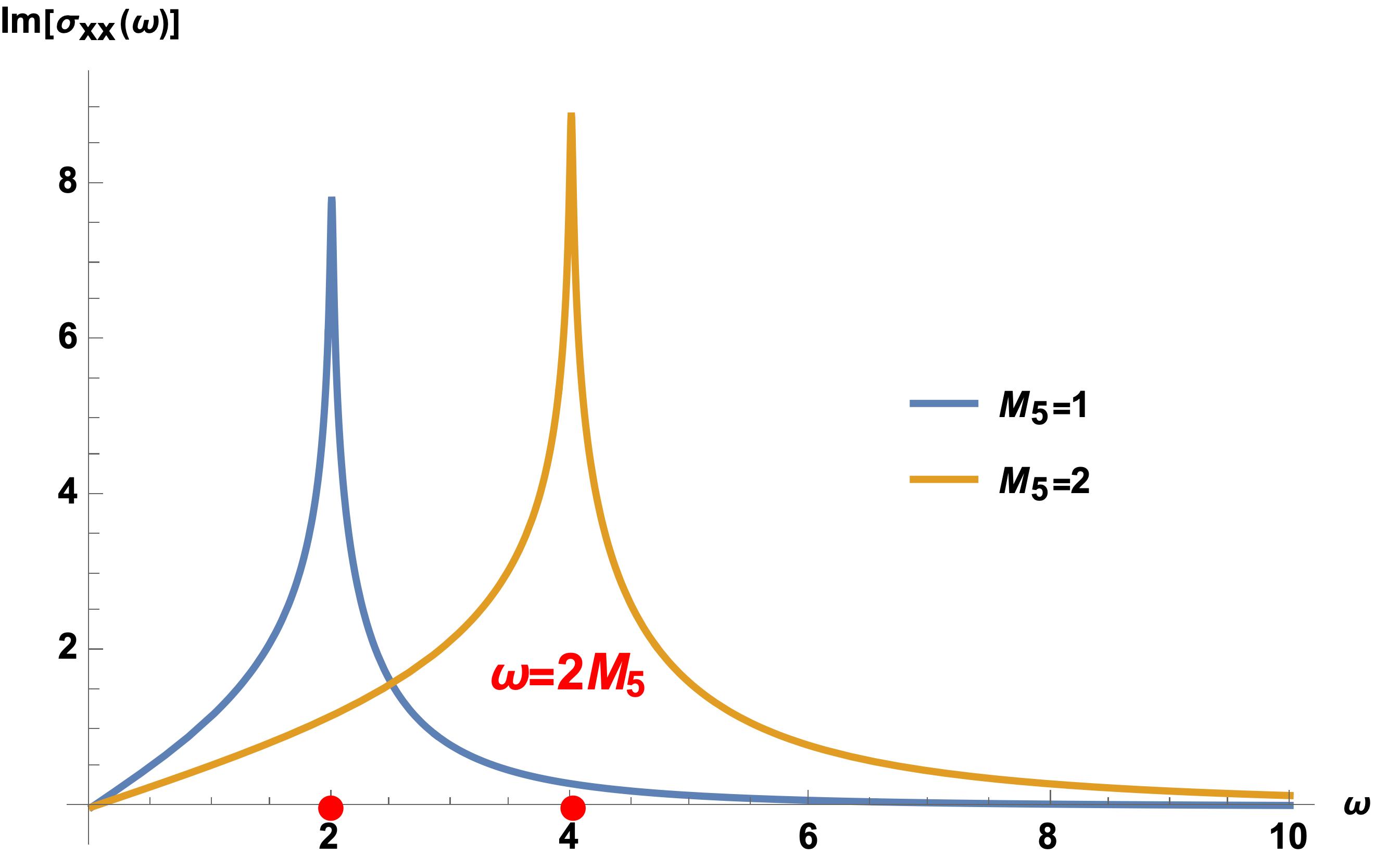}
			\caption{$Im[\sigma_{xx}(\omega)]$}
		\end{subfigure}
		\caption{Transport for pseudoscalar interaction in holographic setup}
		\label{holocon}
	\end{figure}

	From the spectral function (Figure \ref{sfun}), one extract gap energy $\omega_{10}=2\sqrt{k^2+M^2_5}$ which leads to the longitudinal conductivity for the holographic setup
	\begin{eqnarray}
		\sigma_{xx}(\omega) &=&  2 i \int^{\infty}_{0} \frac{kdk}{(2\pi)^2}\int_{0}^{2\pi} d\theta \frac{2\omega\sqrt{k^2+M^2_5}}{4(k^2+M^2_5)-\omega^2} g^{00}_{xx} \\ 
		&=&   i \int^{\infty}_{0} \frac{kdk}{(2\pi)^2}\int_{0}^{2\pi} d\theta \frac{2\omega\sqrt{k^2+M^2_5}}{4(k^2+M^2_5)-\omega^2} Tr(g^{00}) \nonumber \\
		&=&   i \int^{\infty}_{0} \frac{kdk}{(2\pi)^2}\int_{0}^{2\pi} d\theta \frac{2\omega\sqrt{k^2+M^2_5}}{4(k^2+M^2_5)-\omega^2} \frac{k^2+2M_5^2}{4(k^2+M_5^2)^2} . \nonumber  
	\end{eqnarray}
	Performing the integral we get
	\begin{eqnarray} 
		\sigma_{xx}(\omega) &=& \frac{i}{4\pi}\left[ \left(\frac{1}{4}+\frac{M^2_5}{\omega^2}\right)\log\left(\frac{2|M_5|+\omega}{2|M_5|-\omega}\right)-\frac{|M_5|}{\omega}\right] ~.
	\end{eqnarray}
	This is one of the main results in this paper.  We plot the above  conductivity  in Figure \ref{holocon}. 
	In the low frequency limit, we may write
	\begin{eqnarray}
		\mathcal{G} = \int^{\infty}_{0} \frac{kdk}{(2\pi)^2}\int_{0}^{2\pi} d\theta \frac{1}{2\sqrt{k^2+M^2_5}} \frac{k^2+2M_5^2}{4(k^2+M_5^2)^2} =\frac{1}{12\pi |M_5|} ~~.
	\end{eqnarray}
	At low frequency, holographic conductivity reads
	$$\sigma_{\mu\nu}(\omega) = i \omega  \frac{1}{12\pi|M_5|} \delta_{\mu\nu} -\frac{1}{4\pi}sign(M_5)\epsilon_{\mu\nu} .$$
	The longitudinal conductivity is 
	$$Re[\sigma_{\mu\mu}]=0,\quad  for \quad \omega\leq 2M_5, $$ which shows that the system is in insulating state. Since the longitudinal part has  only imaginary part,    we can write 
	$$\sigma_{xx} = \frac{i\omega}{12\pi |M_5|}$$
	and we know the relation between the $\sigma_{xx}$ and dielectric constant $\kappa$ as we can see  in Appendix \ref{appenD}: 
	\begin{eqnarray}
		{\kappa = 1+ \frac{1}{i\omega\epsilon_0} \sigma_{xx}}
		\label{dielc}. 
	\end{eqnarray}
	Using these two equations, we   obtain the holographic dielectric constant at zero temperature : 
	\begin{eqnarray}
		\kappa = 1+ \frac{1}{12\pi\epsilon_0 |M_5|} ~.
	\end{eqnarray}
   These results coincide with those obtained for two-dimensional Dirac materials, where \(M_5\) can be identified with the Dirac mass-gap parameter \(m\) \cite{QGTRachel}. The agreement between our holographic transport results and the corresponding field-theoretic results for two-dimensional Dirac materials \cite{QGTRachel} can be understood from the semi-topological nature of the QGT.
   
   Although strong coupling broadens the single-particle poles into an incoherent or unparticle-like continuum across the Dirac cone, the low-energy effective Hamiltonian is defined from the static limit of the retarded Green's function,
   $H_T(\mathbf{k})=-\mathbb{G}_R^{-1}(\mathbf{k},\omega=0)= h_0(\mathbf{k})\,\mathbb{I} + \mathbf{h}(\mathbf{k})\cdot \boldsymbol{\sigma}.$ In this static limit, the strong-coupling self-energy enters as an overall scalar renormalization of the bare Green's function, such that $
   \mathbb{G}_R(\mathbf{k},0)\approx l(\mathbf{k})\,\mathbb{G}_0(\mathbf{k},0),
   $ where \(l(\mathbf{k})\) is a scalar interaction-dependent factor. Equivalently, this scalar dressing rescales the vector \(\mathbf{h}(\mathbf{k})\) appearing in the effective Hamiltonian without changing its direction in the internal-state space.
   
   Since the QGT depends only on the normalized eigenvectors $\hat{\mathbf{h}}(\mathbf{k})$, or equivalently on the corresponding unit vector on the Bloch sphere, this overall scalar factor does not affect the quantum geometry. The scalar interaction dressing therefore leaves the orientation of the eigenvectors unchanged and, consequently, does not modify either the quantum metric or the Berry curvature. Upon identifying the holographic parameter $(M_5)$ with the Dirac mass gap $(m)$, the ground-state quantum geometry of the holographic system consequently takes the same mathematical form as that of two-dimensional Dirac materials.
   
   The capacitive and dielectric responses of band insulators are governed by the quantum metric integrated across the interband gap \cite{QGTRachel}. The agreement observed here at $(T=0)$ therefore indicates that these ground-state quantum-geometric responses are insensitive to the scalar strong-coupling dressing, reflecting their semi-topological character.

	\section{Transport at finite temperature}
	We now want to calculate the transport at finite temperature and understand the role of black hole geometry and Fermi-Dirac function. The conductivity can generally be decomposed into inter-band and intra-band contributions, $ \sigma_{xx}(\omega,T)=\sigma_{xx}^{\rm intra}(\omega,T)+ \sigma_{xx}^{\rm inter}(\omega,T).$ The present analysis focuses on the inter-band contribution, which is directly related to the quantum geometric tensor (QGT) and therefore isolates the intrinsic geometric contribution to the optical response. 
	
	In the holographic description, moreover, the system is strongly coupled and does not possess a controlled description in terms of sharply defined weakly interacting quasiparticles. The finite-temperature response is instead encoded in the retarded Green's function and the black-hole geometry, with the horizon providing the thermal scale. Thus, we retain the inter-band contribution associated with the holographic spectral structure and QGT (see Appendix B for detailed justification). 
	
	\subsection{Field theory result }
	We start with the finite-temperature calculation for 2D Dirac materials.  The 
	longitudinal conductivity at finite temperature reads 
	\begin{eqnarray}
		\sigma_{xx}(\omega) &=& i \int_{\text{BZ}}  \frac{\tilde{f}_{01} \omega_{10}\omega}{\omega_{01}^2-\omega^2} Tr[g^{01}]
		\label{52} \\
		&=&  i \int^{\infty}_{0} \frac{kdk}{(2\pi)^2}\int_{0}^{2\pi} d\theta  \frac{2\omega \tilde{f}_{01}\sqrt{k^2+m^2}}{4(k^2+m^2)-\omega^2} \frac{k^2+2m^2}{4(k^2+m^2)^2} ~~~~~
	\end{eqnarray}
	where $$\tilde{f}_{01}= f_0(T)(1-f_1(T)) 
	$$ and $f_n(T)$ is Fermi-Dirac distribution function for $n^{th}$ energy state.
	For zero temperature limit,  $\tilde{f}_{01}= 0$ if both are (un)occupied, while it is 1 if only ground state $E_0$ is occupied.  In high temperature limit, 
	$\tilde{f}_{01}$ saturate  $1/4$. 
	The temperature dependence  of the transport in field theory   is coming from FD distribution function.  
	\\
	
	\begin{figure}[h!]
		\centering
		\includegraphics[scale=0.4]{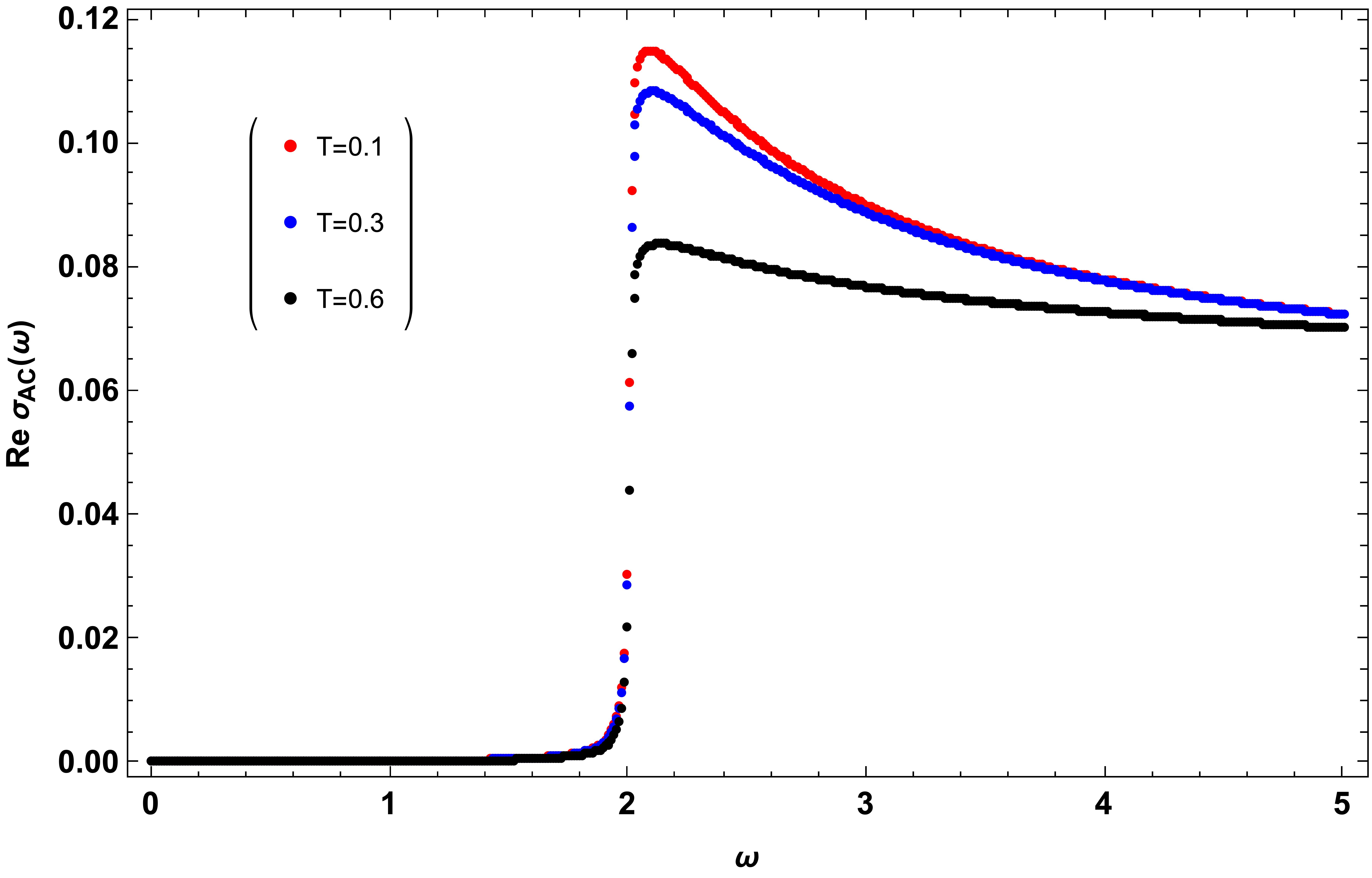}
		\caption{Finite temperature conductivity in 2D Dirac materials from QGT}
		\label{figCH2D}
	\end{figure}
	We   numerically computed the AC-conductivity at different temperature and DC-conductivity  at $m=1$, which are shown in Figure \ref{figCH2D}.  For $\omega\leq 2m$, Figure 3(a)   shows 	that it is in the insulating phase whereas    conductivity is finite  for  $\omega > 2m$.  Note that if  incident frequency is larger than the band gap energy, there is inter-band transition which  results in  insulator-metallic transition and gives finite   conductivity as frequency increases. As temperature goes up, thermal excitation plays crucial role and the peak values decrease at higher temperature because of the collision of particles  caused by thermal excitations. 

	\subsection{Holography results}
	The basic strategy is to generalize the transport formula, Eq.~\eqref{52}, to the holographic framework by utilizing the holographic quantum geometric tensor $g_{\mu\nu}$. However, a subtle conceptual issue arises regarding the definitions of the interband transition frequencies $\omega_{01}$ and the thermal occupation factors $\tilde{f}_{01} = f_0(1 - f_1)$, both of which conventionally rely on discrete single-particle band indices.
	
	In holography, the notion of a two-band structure naturally maps onto distinguishable internal degrees of freedom, such as sublattice pseudospins ($A$ and $B$) or the positive and negative energy branches of the Dirac cone. If necessary, this structure can be generalized to multi-band systems by introducing additional spinor flavor multiplets in the bulk. For a single-flavor bulk model, the two effective branches are operationally identified from the poles of the Green's function, specifically where the real part of the denominator of $\mathbb{G}_R(\mathbf{k}, \omega)$ vanishes. This yields well-defined effective dispersion curves $\omega_0(\mathbf{k})$ and $\omega_1(\mathbf{k})$. Crucially, these excitation energies become intrinsically temperature-dependent already at the classical level in the bulk, $\omega_{nm} = \omega_{nm}(\mathbf{k}, T)$, because the background AdS-Schwarzschild metric directly incorporates the horizon radius $z_h$, which determines the boundary Hawking temperature $T = 3/(4\pi z_h)$. This behavior stands in sharp contrast to non-interacting condensed-matter models, where bare band dispersions are strictly temperature-independent, and thermal effects enter only through occupation functions. In standard quantum field theory, temperature-dependent dispersions emerge only after performing an  RPA resummation of self-energy. In the holographic dual, the classical gravitational field in the black hole background inherently resums these   loop corrections, which is a  genuine  feature of the duality.

	A separate issue concerns the statistical factor \(\tilde f_{01}(T)\) appearing in the conventional finite-temperature Kubo formula. In a strongly coupled holographic system, sharply defined quasiparticles cease to exist, precluding a description based on single-particle states occupying bare energy levels. At zero temperature, a half-filled many-body ground state corresponds to setting $\tilde{f}_{01} = 1$. Even at  finite temperature,  
	we assume that only the ground state is occupied whatever is the temperature. 
	This is because the excited state of many particle system is no more defined by the one particle excitation since there is   no 1-particle concept in strongly correlated system. 
There is no tower of one particle  states  that is available and the ground state of many particle  is defined by the all particles. 
 We therefore adopt the ansatz $\tilde{f}_{01} \rightarrow 1$ for all temperatures in holography to isolate the intrinsic inter-band QGT-controlled response. 
	
	The modified Kubo formula for finite-temperature holographic transport takes the form:
	\begin{equation}
		\sigma_{\mu\nu}(\omega) = -\frac{2 i e^2}{\hbar}\sum_{n\neq m}\int_{\text{BZ}} \frac{\omega_{nm}(T)\omega}{\omega_{nm}(T)^2-\omega^2} g_{\mu\nu}^{nm} -\frac{e^2}{\hbar}\sum_{n\neq m}\int_{\text{BZ}} \frac{\omega_{nm}(T)^2}{\omega_{nm}(T)^2-\omega^2} F_{\mu\nu}^{nm}.
		\label{condM1}
	\end{equation}
	To systematically evaluate this proposal, we compute the transport properties without modifying Eq.~(\ref{condequ1}) for this holographic system and compared the results from both formulas, Eq.~(\ref{condequ1}) and Eq.~(\ref{condM1}).\\
	
\begin{figure}[h!]
		\centering
		\begin{subfigure}[b]{0.3\textwidth}
			\centering
			\includegraphics[scale=0.30]{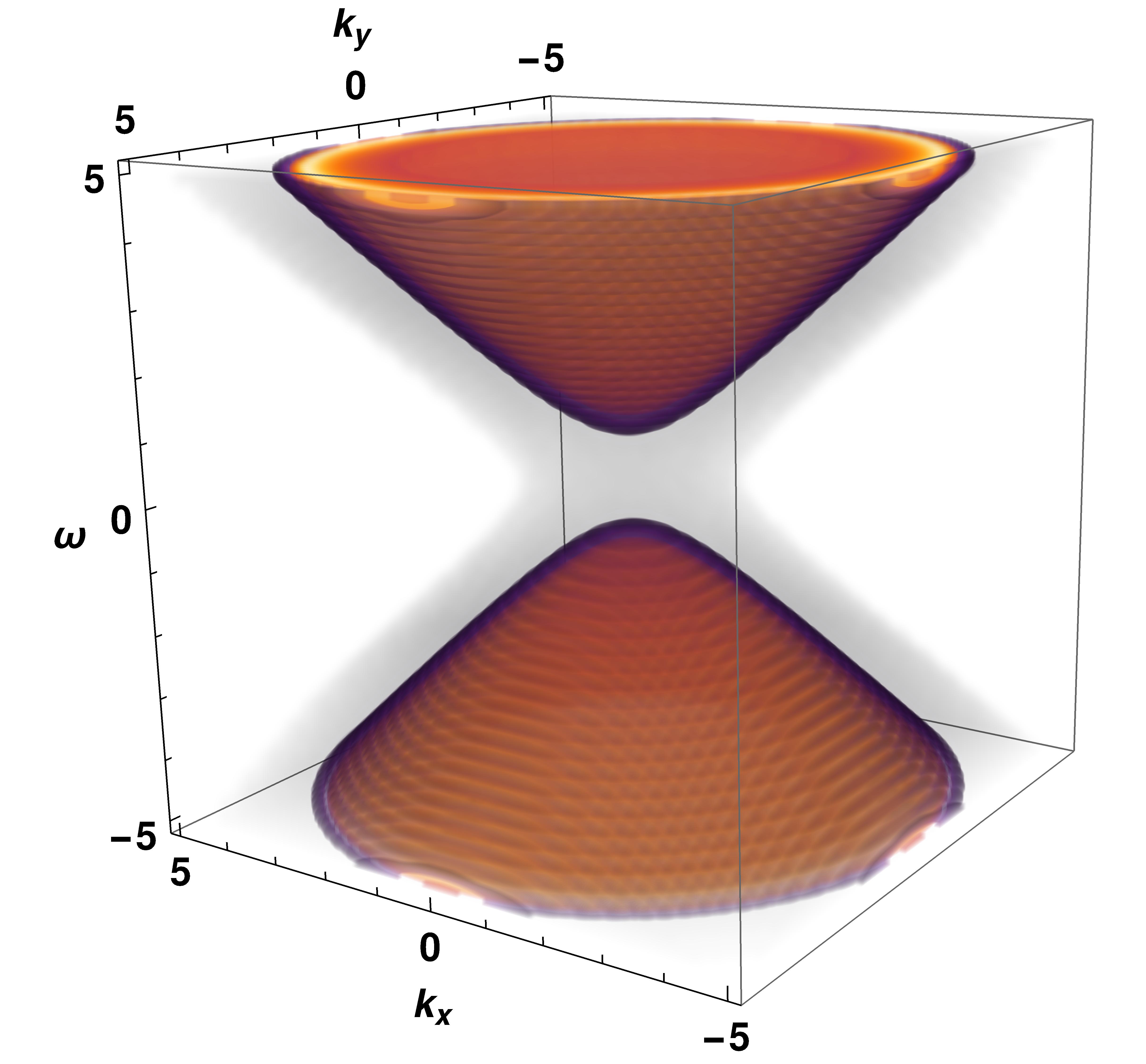}
			\caption{T=0.1}
		\end{subfigure}
		\hfil
		\begin{subfigure}[b]{0.3\textwidth}
			\centering
			\includegraphics[scale=0.30]{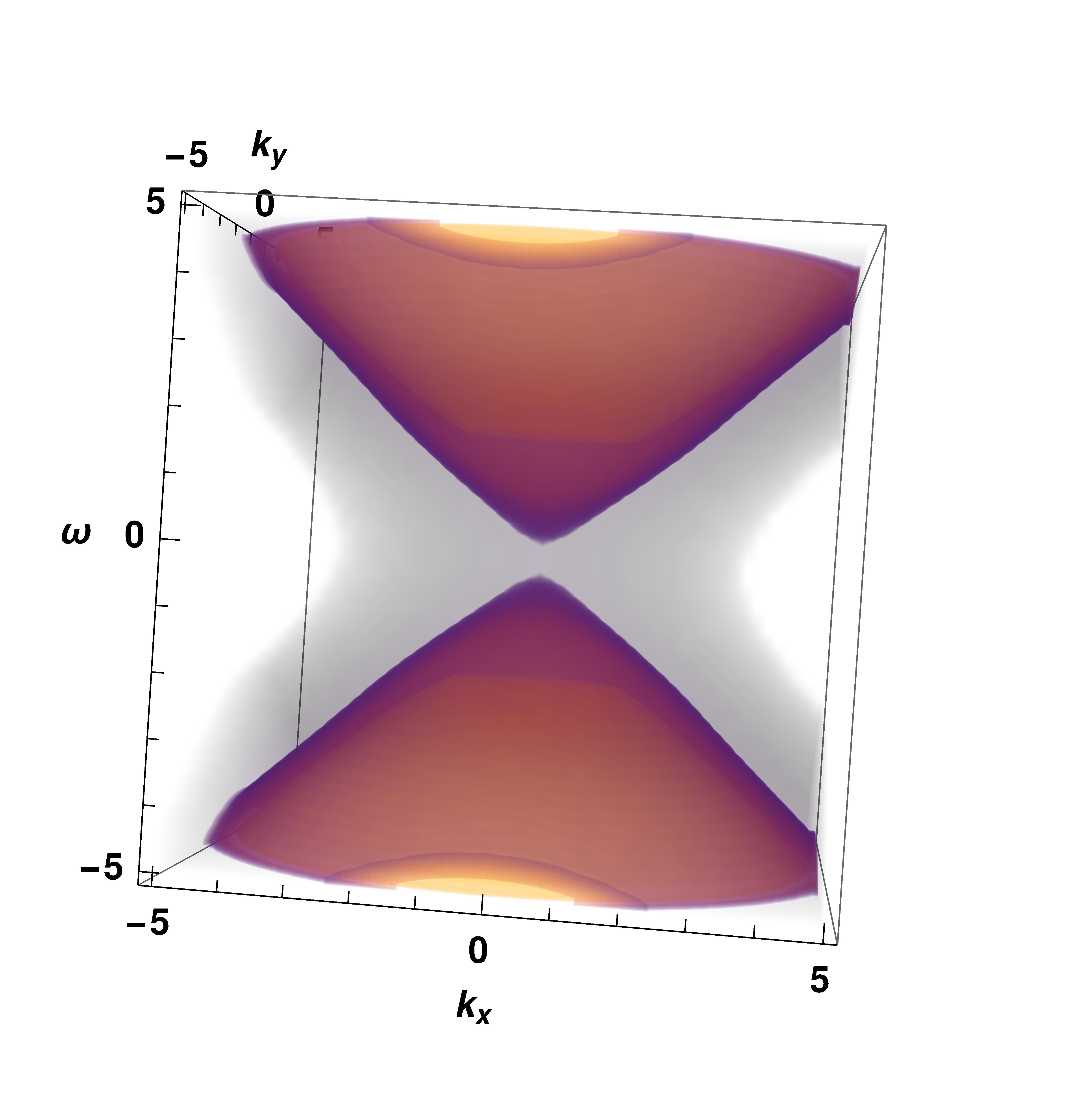}
			\caption{T=0.4}
		\end{subfigure}
		\hfil
		\begin{subfigure}[b]{0.3\textwidth}
			\centering
			\includegraphics[scale=0.30]{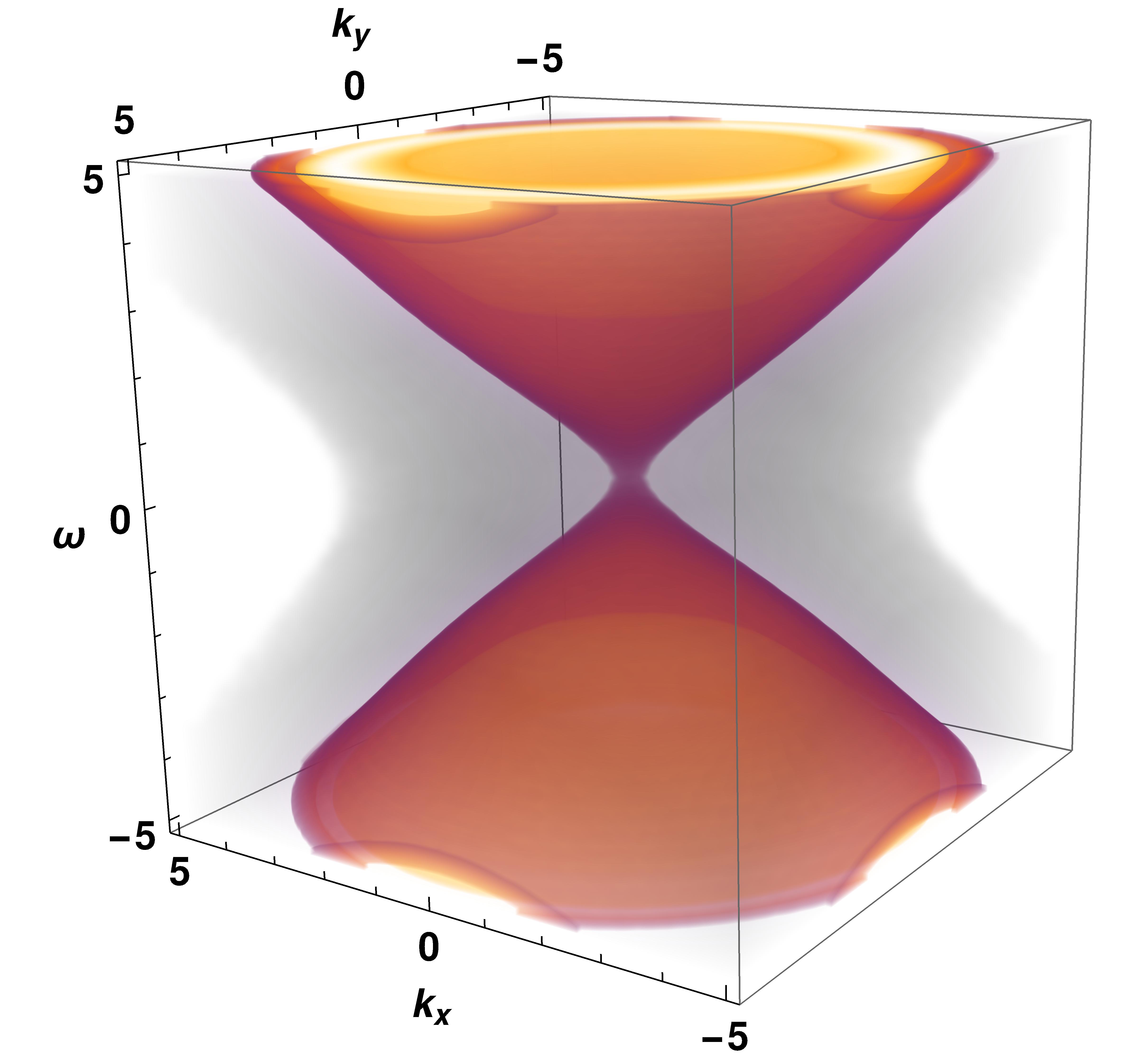}
			\caption{T=0.5}
		\end{subfigure}
		\caption{Fermionic spectral function at different temperatures for fixed $M_5=1$}
		\label{figSF3D}
	\end{figure}

\begin{figure}[h!]
		\centering
		\includegraphics[scale=0.30]{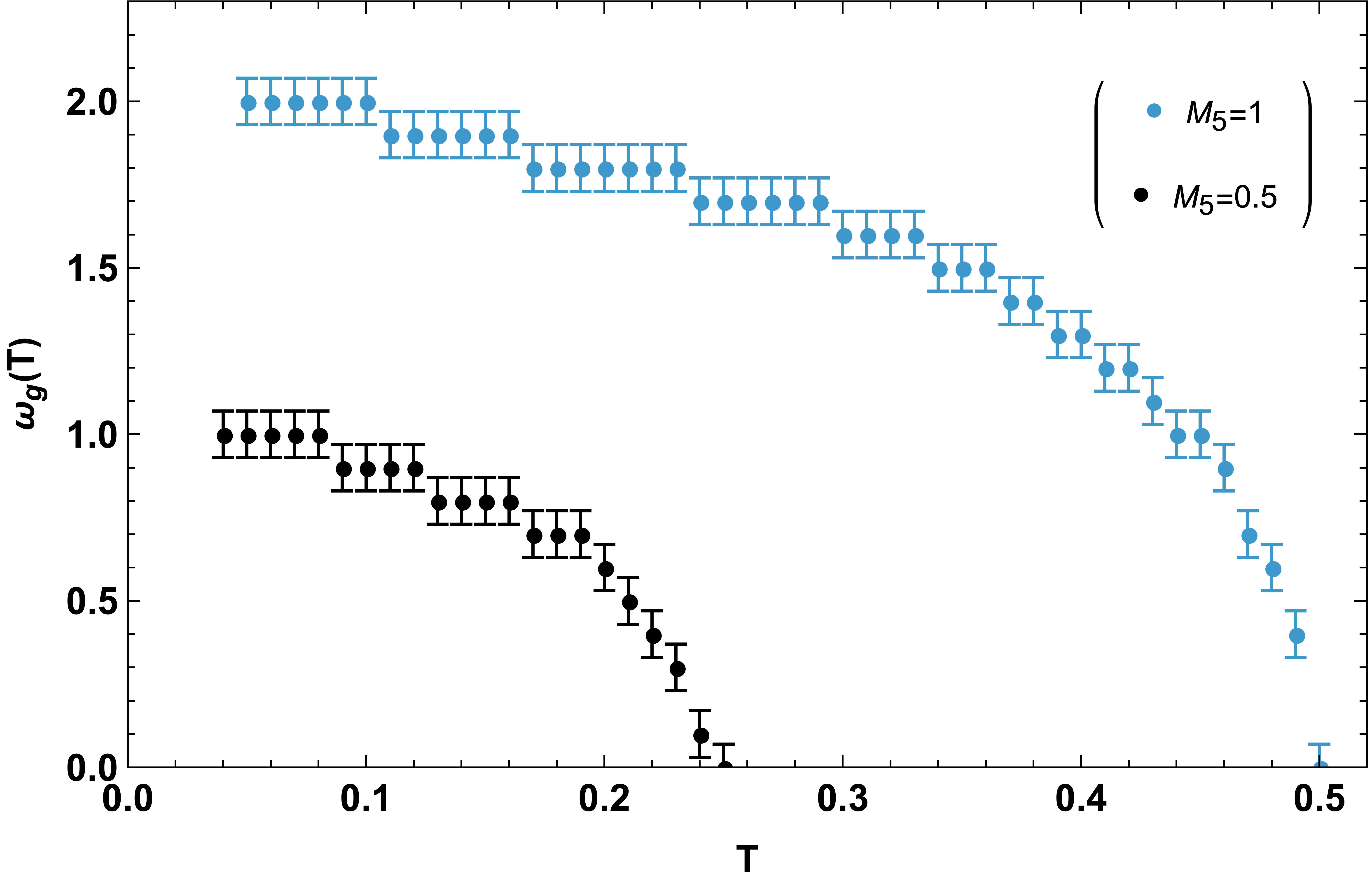}
		\caption{Energy gap as function temperature for different values of $M_5$; $T_c=0.5M_5$}
		\label{gap1}
	\end{figure}

	In our previous paper \cite{HoloTopo}, we reported that the QGT is temperature independent in holography, even when the background is a black-hole geometry. In the Kubo formula above, the temperature dependence enters only through the energy gaps $\omega_{nm}$. By contrast, in conventional condensed-matter treatments $\omega_{nm}$ is temperature independent, and the temperature dependence arises from the Fermi-Dirac (FD) distribution. We now compute the longitudinal conductivity (setting $e^{2}/\hbar = 1$), which takes the form
	\begin{eqnarray}
		\sigma^{xx}(\omega) = i \int_{0}^{\infty} \frac{kdk}{(2\pi)^2} \int_{0}^{2\pi} d\theta \frac{\omega_{10}\omega}{\omega_{10}^2-\omega^2} Tr[g^{01}]
	\end{eqnarray}
	For an AdS-Schwarzschild black-hole background, Fig.~\ref{figSF3D} (fermionic spectral function 3D-plot) shows that the energy gap decreases as the temperature increases which is one of main differences from field theory.  We have shown the temperature dependent frequency gap (in the spectral function) in Figure \ref{gap1}.
	\\
	
\begin{figure}[h!]
		\centering
		\begin{subfigure}[b]{0.49\textwidth}
			\centering
			\includegraphics[scale=0.27]{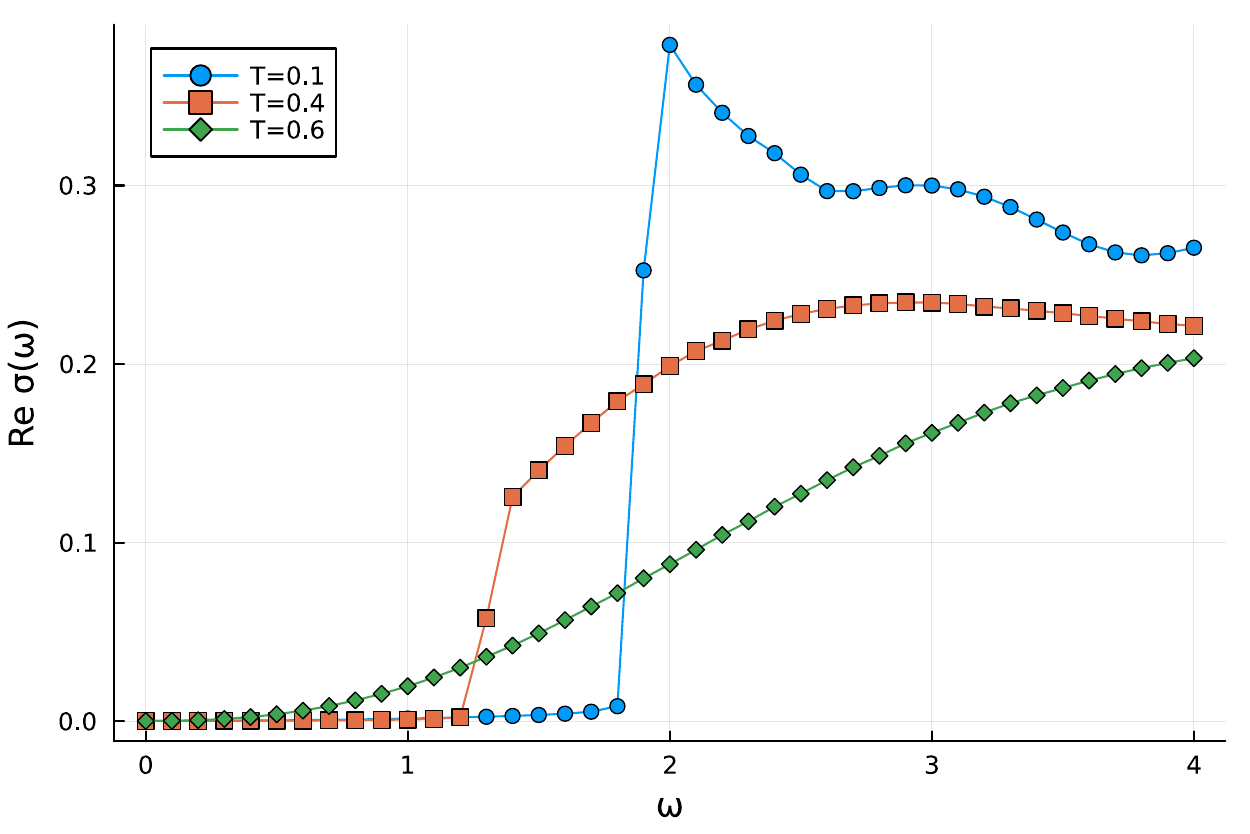}
			\caption{Holographic $\sigma_{AC}^{xx}(\omega)$ with FD}
		\end{subfigure}
		\hfil
		\begin{subfigure}[b]{0.49\textwidth}
			\centering
			\includegraphics[scale=0.27]{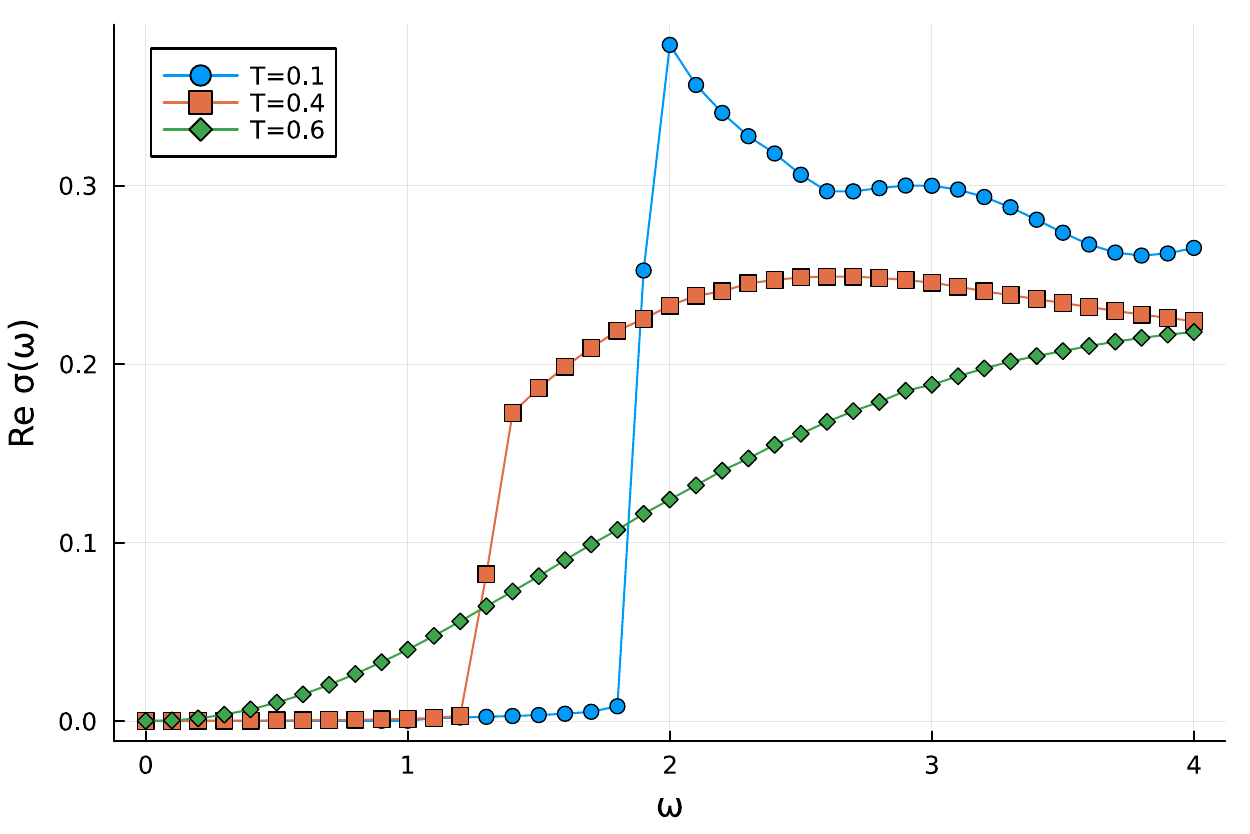}
			\caption{Holographic $\sigma_{AC}^{xx}(\omega)$ without FD}
		\end{subfigure}
		\caption{AC conductivities at various temperatures (FD=Fermi-Dirac function)}
		\label{figAC-Com}
	\end{figure}

	We compute AC conductivity in holography with/without FD distribution function which is shown in Figure \ref{figAC-Com}. No qualitative differences are observed in the holographic AC conductivity, regardless of whether the Fermi-Dirac distribution is included.  As the frequency gap in the fermionic spectral function decreases with increasing temperature, the threshold frequency at which the conductivity changes from vanishing to finite values is correspondingly shifted to lower frequencies. This behavior of the AC conductivity is fully consistent with the temperature dependence of the fermionic spectral function, demonstrating a clear correspondence between the transport and spectral properties of the system. 
	
	While the FD distribution traditionally suppresses AC conductivity in field theories, holographic models replicate this temperature-dependent suppression without explicitly incorporating an FD function. This suggests that the black-hole geometry incorporates thermal effects that, in a weakly coupled description, would conventionally be represented by Fermi--Dirac occupation factors. Furthermore, as temperature increases, the transition frequency decreases, a feature reflected in the fermionic spectral function that contrasts with conventional band theory, where the bandgap remains temperature-independent.
	
	Another observation from holographic transport (Fig.~\ref{figAC-Com}) is that, as the temperature increases toward and beyond \(T_c=0.5M_5\), the frequency-dependent conductivity evolves continuously, rather than exhibiting a sharp jump at the critical frequency corresponding to the gap energy. This behavior reflects the progressive closing of the holographically generated fermionic gap and the corresponding transfer of spectral weight toward low frequencies. To understand the origin and significance of this critical temperature, we first examine the temperature-dependent DC conductivity in the field-theory description and then generalize the analysis to the holographic case, focusing on the present holographic insulating system.
	
	Let us now examine the DC conductivity and its behavior across the temperature-driven insulator--metal crossover. Before turning to the holographic result, it is useful to clarify the definition of the DC conductivity in field theory for a gapped insulating system. The DC conductivity discussed below corresponds to the regularized inter-band contribution considered in this work and should not be interpreted as the complete finite-temperature conductivity including a conventional intra-band Drude component.
	
	In the strictly dissipationless limit, setting \(\omega=0\) in the inter-band Kubo formula gives \(\sigma_{\mathrm{DC}}=0\). Introducing the retarded prescription \(\omega\rightarrow\omega+i\epsilon\), however, yields a finite regularized zero-frequency response. The resulting \(\sigma_{\mathrm{DC}}(T)\) depends on the broadening parameter \(\epsilon\). Importantly, the qualitative temperature dependence is insensitive to the choice of \(\epsilon\): \(d\sigma_{\mathrm{DC}}/dT>0\) in the low-temperature insulating regime, followed by a change of sign at the crossover temperature, while \(d\sigma_{\mathrm{DC}}/dT<0\) in the metallic regime. Thus, \(\epsilon\) serves only as a phenomenological regularization of the zero-frequency response, whereas the insulating-to-metallic crossover is identified from the robust thermal evolution of the conductivity.
	
	In the holographic description, the same crossover is encoded directly in the temperature dependence of the retarded boundary conductivity. In the low-temperature insulating regime, the lowest characteristic excitation scale satisfies \(\omega_{01}\neq0\), and the longitudinal conductivity therefore vanishes in the strict \(\omega\rightarrow0\) limit. A finite zero-frequency response can nevertheless be obtained by evaluating the retarded Green's function at \(\omega+i\epsilon\):
	$
	\sigma_{\mathrm{DC}}(T)=\lim_{\omega\rightarrow0}\mathrm{Re}\,\sigma_{xx}(\omega+i\epsilon),
	$
	where \(\epsilon\) phenomenologically represents finite spectral broadening or a relaxation scale, with \(\tau\sim\hbar/\epsilon\). Although the absolute value of \(\sigma_{\mathrm{DC}}\) consequently depends on \(\epsilon\), its temperature dependence provides a robust diagnostic of the crossover. Specifically, we find \(d\sigma_{\mathrm{DC}}/dT>0\) below \(T_c=0.5M_5\), where the conductivity increases with temperature as thermal effects progressively reduce the gap, characteristic of the insulating regime. At \(T_c\), the temperature derivative changes sign, while for \(T>T_c\) we find \(d\sigma_{\mathrm{DC}}/dT<0\), indicating the metallic regime, in which the low-frequency spectral weight is enhanced and the conductivity decreases with increasing temperature. Thus, within the holographic framework, the sign change of \(d\sigma_{\mathrm{DC}}/dT\), together with the evolution of the fermionic gap and the AC conductivity, provides a consistent characterization of the temperature-driven insulator--metal crossover.

\section{Discussion}

In conclusion, we have developed a unified framework for computing holographic transport properties directly from fermionic Green's functions, thereby avoiding the technically involved bulk one-loop and effective-vertex calculations. The framework provides a common description of spectral, transport, topological, and quantum-geometric properties within a single formalism. Using the effective Hamiltonian constructed from the boundary retarded Green's function, we analyze the ground-state structure and evaluate the interband quantum geometric tensor (QGT) in the gapped phase. We then use the QGT to compute the interband conductivity of the holographic gapped system.

At zero temperature, we calculate the dielectric response using the QGT formalism. Upon identifying the holographic parameter \(M_5\) with the Dirac mass gap \(m\), we find that the ground-state quantum geometry takes the same mathematical form as that of two-dimensional Dirac materials. This agreement can be understood from the semi-topological nature of the QGT: the strong-coupling interaction enters the static Green's function as an overall scalar dressing, which does not change the orientation of the eigenvectors and therefore leaves the quantum metric and Berry curvature unchanged. Consequently, the corresponding ground-state quantum-geometric responses, including the capacitive and dielectric responses, retain the same form as in the corresponding Dirac system.

Our results also reveal distinct quantum and thermal signatures in the holographic transport response. At zero temperature, the system exhibits a frequency-dependent onset of interband conductivity associated with the finite excitation gap. At finite temperature, the AdS--Schwarzschild black-hole background introduces a temperature dependence into the fermionic spectral function and excitation gap. As the temperature increases, the spectral gap progressively closes, while the AC conductivity evolves continuously from the sharp gapped behavior characteristic of the zero-temperature limit to a smoother low-frequency response. This evolution signals the gradual redistribution of spectral weight and provides a direct connection between the closing of the holographic fermionic gap and the temperature-driven insulator--metal crossover.

For a gapped insulator, the strict zero-frequency interband conductivity vanishes in the field-theory description. A finite regularized \(\sigma_{\mathrm{DC}}\) can instead be obtained by employing the retarded prescription \(\omega\rightarrow\omega+i\epsilon\), where \(\epsilon\) represents a phenomenological broadening or relaxation scale. Applying the same prescription to the holographic conductivity, we find \(d\sigma_{\mathrm{DC}}/dT>0\) below \(T_c=0.5M_5\), while the trend reverses to \(d\sigma_{\mathrm{DC}}/dT<0\) above \(T_c\) as the spectral gap closes. The sign change of \(d\sigma_{\mathrm{DC}}/dT\), together with the evolution of the spectral gap and AC conductivity, therefore provides a consistent transport signature of the holographic insulator--metal crossover.

Since the present calculation focuses on the interband quantum-geometric contribution and does not explicitly include thermally activated intraband transport, the resulting DC conductivity should be regarded as a regularized transport diagnostic of the crossover rather than as a complete microscopic calculation of the finite-temperature conductivity.

An important outcome of this work is that thermal effects that would conventionally be described through Fermi--Dirac occupation factors in a weakly coupled theory are instead encoded in the thermal black-hole geometry and the resulting temperature-dependent holographic Green's function. In this sense, the holographic description incorporates thermal many-body effects without requiring an additional occupation prescription based on well-defined single-particle states. This provides a natural framework for studying strongly coupled transport and related quantum-geometric responses, such as dielectric response and capacitance, within a unified setting. The formalism developed here can be generalized to other holographic interactions and strongly correlated systems, opening several directions for future investigation. In particular, extending the present approach to flat-band systems and exploring their transport and quantum-geometric properties within holography would be especially interesting.

	\appendix
	\section{Quantum geometric tensor} 
	\label{a2}
	The gauge-covariant derivative on a quantum state is defined by 
	\begin{align}
		D_{\mu} = \partial_{\mu} + i A_{\mu}
	\end{align}
	where $A_{\mu}=i \langle\psi|\partial_{\mu}\psi\rangle$. The gauge invariant quantity QGT is defined by
	\begin{align}
		Q_{\mu\nu} &\ = \langle D_{\mu}\psi|D_{\nu}\psi\rangle \nonumber \\
		&\ = \langle \partial_{\mu}\psi|\partial_{\nu}\psi\rangle - A_{\mu}A_{\nu}
	\end{align}
	This can be written as $Q_{\mu\nu}=g_{\mu\nu} + \sigma_{\mu\nu}$, where $$g_{\mu\nu}= Re. \langle \partial_{\mu}\psi|\partial_{\nu}\psi\rangle -A_{\mu}A_{\nu} \quad \text{and}\quad \sigma_{\mu\nu}= Im.\langle \partial_{\mu}\psi|\partial_{\nu}\psi\rangle$$ and $A_{\mu}$ is real. $A_{\mu}$ is called Berry connection. Using this Berry connection, we can calculate the Berry curvature as
	\begin{align}
		F_{\mu\nu}  &\ = \partial_{\mu}A_{\nu} -\partial_{\nu}A_{\mu} \nonumber \\
		&\ = 2 i Im.\langle \partial_{\mu}\psi|\partial_{\nu}\psi\rangle
	\end{align}
	Therefore $\sigma_{\mu\nu}=-\frac{i}{2}F_{\mu\nu}$ which leads to
	\begin{align}
		Q_{\mu\nu}=g_{\mu\nu} -\frac{i}{2}F_{\mu\nu}
	\end{align}
	\subsection{General two band systems}
	\label{Ase1}
	Any two by two Hamiltonian can be expressed as 
	\begin{align}
		H = h_0(\mathbf{k}) \, \mathbb{I}_2 + \mathbf{h}(\mathbf{k}) \cdot \boldsymbol{\sigma},
	\end{align}
	where $\mathbb{I}_2$ is the $2 \times 2$ identity matrix, $\mathbf{k} = (k_x, k_y)$, 
	$\mathbf{h} = \big(h_1(\mathbf{k}), h_2(\mathbf{k}), h_3(\mathbf{k})\big)$, 
	$\boldsymbol{\sigma} = (\sigma_1, \sigma_2, \sigma_3)$, and the $\sigma_i$'s are the Pauli matrices. 
	With the normal vector $\hat{\mathbf{h}} = \frac{\mathbf{h}}{|\mathbf{h}|}$, the normalized eigenvectors for this generalized Hamiltonian reads
	\begin{align}
		|\psi_0 \rangle =  \frac{1}{\sqrt{2(1+\hat{h}_3)}}\left( \begin{array}{c}
			-(\hat{h}_1- i \hat{h}_2) \\
			1+\hat{h}_3
		\end{array}\right) && \quad \text{with eigenvalue} \quad h_0- |\mathbf{h}| \\
		|\psi_1 \rangle =  \frac{1}{\sqrt{2(1-\hat{h}_3)}}\left( \begin{array}{c}
			\hat{h}_1- i \hat{h}_2 \\
			1-\hat{h}_3
		\end{array}\right) && \quad \text{with eigenvalue} \quad h_0+ |\mathbf{h}| ~.
	\end{align}
	The Berry connection for the ground state reads
	\begin{align}
		A_{\mu} = i \langle \psi_0 | \partial_{\mu}\psi_0 \rangle = \frac{\hat{h}_1\partial_{\mu}\hat{h}_2-\hat{h}_2\partial_{\mu}\hat{h}_1}{2(1+\hat{h}_3)} ~.
	\end{align}
	The Berry curvature reads
	\begin{eqnarray}
		F_{\mu\nu} &=& \partial_{\mu}A_{\nu} - \partial_{\nu}A_{\mu} \nonumber \\
		&=& \frac{1}{2(1+\hat{h}_3)^2}\left[\hat{h}_1\left(\partial_{\mu}\hat{h}_2\partial_{\nu}\hat{h}_3 -\partial_{\nu}\hat{h}_2\partial_{\mu}\hat{h}_3\right)+\hat{h}_2\left(\partial_{\mu}\hat{h}_3\partial_{\nu}\hat{h}_1 -\partial_{\nu}\hat{h}_3\partial_{\mu}\hat{h}_3\right)+2(1+\hat{h}_3)a_{\mu\nu}^{(12)}\right] \nonumber \\
	\end{eqnarray}
	where ~ $a_{\mu\nu}^{(12)}= \partial_{\mu}\hat{h}_1\partial_{\nu}\hat{h}_2 -\partial_{\nu}\hat{h}_1\partial_{\mu}\hat{h}_2$. After simplification, we obtain
	\begin{eqnarray}
		F_{\mu\nu} &=& \frac{1}{2}\left[\hat{h}_1\left(\partial_{\mu}\hat{h}_2\partial_{\nu}\hat{h}_3 -\partial_{\nu}\hat{h}_2\partial_{\mu}\hat{h}_3\right)+\hat{h}_2\left(\partial_{\mu}\hat{h}_3\partial_{\nu}\hat{h}_1 -\partial_{\nu}\hat{h}_3\partial_{\mu}\hat{h}_3\right)+ \hat{h}_3 a_{\mu\nu} \right]~~~ \\
		&=&\frac{a_{\mu\nu}^{(12)}}{2\hat{h}_{3}} = \frac{a_{\mu\nu}^{(23)}}{2\hat{h}_{1}}=\frac{a_{\mu\nu}^{(31)}}{2\hat{h}_{2}}
	\end{eqnarray}
	where we have used $\sum_{i=1}^{3}\hat{h}_i^2=1$ and $\sum_{i=1}^{3}\hat{h}_{i}\partial_{\alpha}\hat{h}_i=0$. 
	The QGT for the ground state reads
	\begin{align}
		Q_{\mu\nu} = \langle\partial_{\mu}\psi_0|\partial_{\nu}\psi_0\rangle - A_{\mu}A_{\nu} =  \frac{1}{4} \partial_{\mu}\hat{\mathbf{h}}.\partial_{\nu}\hat{\mathbf{h}} - \frac{i}{4} \hat{\mathbf{h}}.\partial_{\mu}\hat{\mathbf{h}}\times \partial_{\nu}\hat{\mathbf{h}}
	\end{align}
	Therefore, the quantum metric and Berry curvature read
	\begin{align}
		g_{\mu\nu} = \frac{1}{4}\partial_{\mu}\hat{\mathbf{h}}.\partial_{\nu}\hat{\mathbf{h}} \quad \text{and} \quad F_{\mu\nu}= \frac{1}{2} \hat{\mathbf{h}}.\partial_{\mu}\hat{\mathbf{h}}\times \partial_{\nu}\hat{\mathbf{h}}=\frac{1}{2\hat{h}_3}\left(\partial_{\mu}\hat{h}_1 \partial_{\nu}\hat{h}_2 - \partial_{\nu}\hat{h}_1 \partial_{\mu}\hat{h}_2 \right) 
	\end{align}
	The above formula is true for any two band systems. We can label the momentum space coordinate index as $\mu=\{k_x,k_y,..\}$.
	
	\section{Derivation of Kubo formula in terms of QGT}
	\label{a3}
	We start from the Kubo formula for the conductivity tensor:
	\begin{equation}
		\sigma^{\mu\nu}(\omega) = \frac{i \bar{n} e^2}{m \omega_+}\delta^{\mu\nu} 
		+ \frac{1}{\hbar \omega_+ A}\int_0^\infty dt \, e^{i\omega_+ t} 
		\langle [ j^\mu(t), j^\nu(0) ] \rangle,
	\end{equation}
	where $\omega_+ = \omega + i\epsilon$, $\bar{n}$ is the carrier density, $A$ is the system area, and $j^\mu$ is the current operator. 
	Let $\{|n\mathbf{k}\rangle\}$ be eigenstates of the Hamiltonian:
	\begin{equation}
		H|n\mathbf{k}\rangle = E_n(\mathbf{k}) |n\mathbf{k}\rangle.
	\end{equation}
	The expectation value of the commutator in thermal equilibrium is
	\begin{equation}
		\langle [j^\mu(t), j^\nu(0)] \rangle 
		= \mathrm{Tr}\left( \rho [j^\mu(t), j^\nu(0)] \right),
	\end{equation}
	Diagonal elements of $\rho$ in the eigenbasis are the Fermi--Dirac distribution:
	\begin{eqnarray}
		\langle n\mathbf{k}|\rho|n\mathbf{k}\rangle = f_n(\mathbf{k}) 
		= \frac{1}{e^{\beta E_n(\mathbf{k})}+1}.
	\end{eqnarray}
	The commutator can be expanded as
	\begin{eqnarray}
		\langle [ j^\mu(t), j^\nu(0)] \rangle 
		&=& \sum_n f_n(\mathbf{k}) \langle n\mathbf{k} | j^\mu(t) j^\nu(0) - j^\nu(0) j^\mu(t) | n\mathbf{k} \rangle \\
		&=& \sum_{nm} f_n(\mathbf{k}) \left( \langle n\mathbf{k} | j^\mu(t) | m\mathbf{k}\rangle \langle m\mathbf{k} | j^\nu(0) | n\mathbf{k}\rangle - \mu \leftrightarrow \nu \right) ~.
	\end{eqnarray}
	In the Heisenberg picture
	$
	j^\mu(t) = e^{iHt/\hbar} j^\mu e^{-iHt/\hbar}
	$, which leads to
	\begin{equation}
		\langle n\mathbf{k} | j^\mu(t) | m\mathbf{k}\rangle 
		= e^{i(E_n - E_m)t/\hbar} \langle n\mathbf{k}| j^\mu | m\mathbf{k}\rangle 
		= e^{i\omega_{nm}t} j^\mu_{nm}(\mathbf{k}),
	\end{equation}
	with $\omega_{nm} = (E_n-E_m)/\hbar$.
	Thus,
	\begin{align}
		\langle [ j^\mu(t), j^\nu(0)] \rangle 
		&= \sum_{nm} f_n(\mathbf{k}) \left( e^{i\omega_{nm}t} j^\mu_{nm} j^\nu_{mn} - e^{-i\omega_{nm}t} j^\nu_{nm} j^\mu_{mn} \right).
	\end{align}
	Relabel $n\leftrightarrow m$ in the second term, giving
	\begin{equation}
		\langle [ j^\mu(t), j^\nu(0)] \rangle 
		= \sum_{nm} f_{nm}(\mathbf{k}) j^\mu_{nm}(\mathbf{k}) j^\nu_{mn}(\mathbf{k}) e^{i\omega_{nm}t},
	\end{equation}
	where $f_{nm} = f_n - f_m$.
	Finally, including the momentum sum,
	\begin{equation}
		\langle [ j^\mu(t), j^\nu(0)] \rangle 
		= \sum_{nm}\int_{\text{BZ}} \frac{A\, d^dk}{(2\pi)^d} f_{nm}(\mathbf{k}) j^\mu_{nm}(\mathbf{k}) j^\nu_{mn}(\mathbf{k}) e^{i\omega_{nm}(\mathbf{k})t}.
	\end{equation}
	Substitute the above into conductivity expression:
	\begin{align}
		\sigma^{\mu\nu}(\omega) 
		&= \frac{i \bar{n} e^2}{m \omega_+}\delta^{\mu\nu} 
		+ \frac{1}{\hbar \omega_+}\sum_{nm}\int_{\text{BZ}} \frac{d^dk}{(2\pi)^d} f_{nm} j^\mu_{nm} j^\nu_{mn} \int_0^\infty dt \, e^{i(\omega_{nm}+\omega_+)t}.
	\end{align}
	Thus,
	\begin{equation}
		\sigma^{\mu\nu}(\omega) = \frac{i \bar{n} e^2}{m \omega_+}\delta^{\mu\nu} 
		+ \frac{i}{\hbar \omega_+}\sum_{nm}\int_{\text{BZ}} \frac{d^dk}{(2\pi)^d} \frac{ f_{nm} j^\mu_{nm} j^\nu_{mn} }{\omega_{nm}+\omega_+}.
	\end{equation}
	Notice the prefactor:
	\begin{equation}
		\frac{1}{\omega_+(\omega_{nm}+\omega_+)}.
	\end{equation}
	Use the identity:
	\begin{equation}
		\frac{1}{\omega_+(\omega_{nm}+\omega_+)} 
		= \frac{1}{\omega_{nm}} \left( \frac{1}{\omega_+} - \frac{1}{\omega_{nm}+\omega_+} \right).
	\end{equation}
	The term proportional to $1/\omega_+$ cancels exactly with the diamagnetic contribution $\tfrac{i\bar{n}e^2}{m\omega_+}\delta^{\mu\nu}$. This condition leads to define the ratio between charge density and mass of the carriers in following way:
	\begin{eqnarray}
		\frac{\bar{n}}{m}=-\frac{1}{\hbar e^2} \sum_{n m} \int_{\text{BZ}} \frac{f_{nm}}{\omega_{nm}}  j^\mu_{nm} j^\mu_{mn}
	\end{eqnarray}
	Thus the final expression of conductivity becomes
	\begin{equation}
		\sigma^{\mu\nu}(\omega) = -\frac{i}{\hbar} \sum_{nm}\int_{\text{BZ}} \frac{d^dk}{(2\pi)^d}
		\frac{ f_{nm}(\mathbf{k}) j^\mu_{nm}(\mathbf{k}) j^\nu_{mn}(\mathbf{k}) }{\omega_{mn}(\mathbf{k}) \, [\omega_{mn}(\mathbf{k})+\omega]}.
	\end{equation}
	Since $j^\mu = e v^\mu$, we compute the matrix elements in the Bloch eigenbasis:
	\begin{equation}
		j^\mu_{nm}(\mathbf{k}) = \langle n\mathbf{k}| j^\mu |m\mathbf{k}\rangle 
		= e \langle n\mathbf{k}| v^\mu |m\mathbf{k}\rangle.
	\end{equation}
	Substituting $v^\mu = \tfrac{1}{\hbar}\partial_{\mu} H(\mathbf{k})$, we obtain
	\begin{eqnarray}
		j^\mu_{nm}(\mathbf{k}) &=& \frac{e}{\hbar}\langle n\mathbf{k}| \partial_{\mu} H(\mathbf{k}) |m\mathbf{k}\rangle \\
		&=& \frac{e}{\hbar} \left[ \langle n\mathbf{k}| \partial_{\mu} \left(  H(\mathbf{k}) |m\mathbf{k}\rangle \right) -  \langle n\mathbf{k}| H(\mathbf{k})|\partial_{\mu} m\mathbf{k}\rangle \right]\\
		j^\mu_{nm}(\mathbf{k}) &=& \frac{e}{\hbar}\partial_{\mu}E_{m} \langle nk|mk\rangle + \frac{e}{\hbar} \left(E_{m}-E_{n}\right) \langle n\mathbf{k}|\partial_{\mu} m\mathbf{k}\rangle = e\delta_{nm}\partial_{\mu}\omega_m  -i e \omega_{nm} A_{\mu}^{nm}~~~~~~~~~~
	\end{eqnarray}
	The conductivity becomes
	\begin{eqnarray}
		\sigma^{\mu\nu}(\omega) &=& -\frac{i e^2}{\hbar} \sum_{n\neq m}\int_{\text{BZ}} f_{nm}(\mathbf{k}) \omega_{nm}(\mathbf{k})
		\frac{  A_\mu^{nm}(\mathbf{k}) A_\nu^{mn}(\mathbf{k}) }{ \, [\omega_{nm}(\mathbf{k})+\omega]} -\frac{i e^2}{\hbar} \sum_{n}\int_{\text{BZ}} \frac{f^{\prime}_n(\mathbf{k})}{\omega}\partial_{\mu}\omega_n(\mathbf{k}) \partial_{\nu}\omega_n(\mathbf{k}) \nonumber \\
		&=& \sigma^{\mu\nu}_{\mathrm{intra}}(\omega) + \sigma^{\mu\nu}_{\mathrm{inter}}(\omega)
		\label{condall}
	\end{eqnarray}
	where the total conductivity tensor decomposes into intra-band and inter-band transitions.
	The expression of spectral weight becomes 
	\begin{eqnarray}
		\frac{\bar{n}}{m}=\frac{1}{\hbar} \sum_{n m} \int_{\text{BZ}} f_{nm}\omega_{nm} A_\mu^{nm}(\mathbf{k}) A_\mu^{mn}(\mathbf{k}) ~.
	\end{eqnarray}
	
	For intraband $m=n$, we can substitute $\omega_{nn}=0$ in the above expression. All contributions to the transport properties arise from interband transitions, whereas the intraband contribution vanishes because the derivative of the Fermi function becomes a delta function at zero temperature \cite{QGTRachel}. The interband conductivity reads
	\begin{eqnarray}
		\sigma^{\mu\nu}(\omega) &=& -\frac{i e^2}{\hbar} \sum_{n\neq m}\int_{\text{BZ}} \left(f_n(1-f_m) -f_m(1-f_n) \right) \omega_{nm}(\mathbf{k})
		\frac{  A_\mu^{nm}(\mathbf{k}) A_\nu^{mn}(\mathbf{k}) }{ \, [\omega_{nm}(\mathbf{k})+\omega]} \nonumber \\
		&=& -\frac{i e^2}{\hbar} \sum_{n\neq m}\int_{\text{BZ}} f_n(1-f_m) \omega_{nm} \left( \frac{A_{\mu}^{nm}A_{\nu}^{mn}}{\omega_{nm}+\omega} - \frac{A_{\nu}^{nm}A_{\mu}^{mn}}{\omega_{nm}-\omega} \right) \\
		&=& -\frac{i e^2}{\hbar} \sum_{n\neq m}\int_{\text{BZ}} \tilde{f}_{nm} \omega_{nm} \left( \frac{A_{\mu}^{nm}A_{\nu}^{mn}}{\omega_{nm}+\omega} - \frac{A_{\nu}^{nm}A_{\mu}^{mn}}{\omega_{nm}-\omega} \right)
	\end{eqnarray}
	From the definition of metric and Berry connection
	\begin{align}
		g_{\mu\nu}^{nm} = \frac{1}{2}\left(A_\mu^{nm} A_\nu^{mn} + A_\nu^{nm} A_\mu^{mn}   \right)  \quad 	F_{\mu\nu}^{nm} = i\left(A_\mu^{nm} A_\nu^{mn} - A_\nu^{nm} A_\mu^{mn}   \right)
	\end{align}
	Therefore, the interband conductivity in terms of geometry and topology reads
	\begin{eqnarray}
		\sigma^{\mu\nu}(\omega) &=& -\frac{2 i e^2}{\hbar}\sum_{n\neq m}\int_{\text{BZ}}  \frac{\tilde{f}_{nm} \omega_{nm}\omega}{\omega_{nm}^2-\omega^2} g_{\mu\nu}^{nm} -\frac{e^2}{\hbar}\sum_{n\neq m}\int_{\text{BZ}}  \frac{\tilde{f}_{nm} \omega_{nm}^2}{\omega_{nm}^2-\omega^2} F_{\mu\nu}^{nm}
	\end{eqnarray}
	The spectral weight yields
	\begin{eqnarray}
		\frac{\bar{n}}{m}=\frac{1}{\hbar} \sum_{n\neq m} \int_{\text{BZ}} f_{nm}\omega_{nm}  g_{\mu\mu}^{mn}(\mathbf{k}) ~~.
	\end{eqnarray}
	which is consistent with the corresponding results reported in the literature \cite{QGTRachel}.\\
	
In both the massive 2D Dirac material and the holographic setup, we restrict our finite-temperature transport calculations exclusively to interband transitions. The focus on the inter-band response is motivated by the quantum-geometric description of insulating systems, in which the reactive longitudinal conductivity is directly related to the quantum metric and, consequently, to the dielectric and capacitive response.

In standard band theory, the intra-band (Drude) weight is governed by the derivative of the Fermi distribution:
\begin{equation}
	\sigma^{\mu\nu}_{\mathrm{intra}}(\omega) \propto
	\frac{i}{\omega + i\epsilon}
	\sum_n \int_{\mathrm{BZ}} d^2k
	\left(-\frac{\partial f_n}{\partial E}\right)
	\partial_{\mu}E_n(\mathbf{k})
	\partial_{\nu}E_n(\mathbf{k}).
\end{equation}
In our present framework, we focus exclusively on the inter-band contribution $\sigma^{\mu\nu}_{\mathrm{inter}}(\omega)$. This treatment is physically justified by the following considerations.

In an undoped, half-filled insulator, the chemical potential resides in the middle of the energy gap ($\mu = 0$), where the density of states is zero. At zero temperature, the Fermi-surface derivative strictly reduces to a Dirac delta function:
\[
-\frac{\partial f_n}{\partial E}
= \delta(E_n(\mathbf{k}) - E_F) = 0
\quad \forall\ \mathbf{k} \in \mathrm{BZ}.
\]
Consequently, the intra-band contribution vanishes identically at $T = 0$, leaving the inter-band quantum geometric tensor (QGT) as the sole origin of reactive and optical transport.

At finite temperature ($T > 0$), thermal excitation across the gap yields a non-vanishing tail $-\partial f_n / \partial E \sim \mathcal{O}(e^{-\text{gap} / 2T})$, which in conventional weakly coupled semiconductors generates an activated intra-band Drude peak. In this work, neglecting this term is an explicit approximation adopted to isolate the \emph{quantum-geometric response} of the gapped system. In particular, the reactive component of the inter-band conductivity is directly tied to the quantum metric $g_{\mu\nu}$ and determines the intrinsic dielectric constant and quantum capacitance of the insulating state. The temperature-driven insulator--metal transition is captured through the continuous collapse of the holographic spectral gap $\omega_{10}(T)$ rather than an empirical addition of a thermal Drude peak.

In strongly coupled holographic matter, discrete single-particle quasiparticle poles are replaced by an unparticle continuum. Without explicit finite chemical potential (charge doping) or an explicit boundary conservation mechanism leading to a sharp Drude peak, the system lacks long-lived, freely accelerating quasiparticles within an isolated band dispersion. The collective thermal transport across the spectral gap is thus naturally mediated by the interband matrix elements extracted from the effective topological Hamiltonian.	
	

	\section{Dielectric constant}
	\label{appenD}
	The longitudinal conductivity is expressed as \cite{Ge:2016lyn}
	\begin{eqnarray}
		\sigma^{xx} &=& \frac{J_x}{E_x}= \frac{I_x}{A}\frac{d}{V_x}=\frac{d}{A} I_x \frac{C}{Q_x} = C \frac{d}{A} \frac{1}{Q_x}\frac{dQ_x}{dt} = C \frac{d}{A} i\omega \\
		\sigma^{xx} &=& i\omega \frac{dC}{A}
	\end{eqnarray}
	We know that the capacitance of a matter can be written in terms of Dielectric constant $\kappa$ as following:
	\begin{eqnarray}
		C = (\kappa -1)\epsilon_0 \frac{A}{d}
	\end{eqnarray}
	where $\epsilon_0$ is the vacuum permittivity. Then, conductivity reads
	\begin{eqnarray}
		\sigma^{xx} &=& i\omega (\kappa -1)\epsilon_0
	\end{eqnarray}
	In vacuum, $\kappa=1$, then $\sigma^{xx}=0$. We can write dielectric constant in terms of conductivity:
	\begin{eqnarray}
		\boxed{\kappa = 1+ \frac{1}{i\omega\epsilon_0} \sigma^{xx}}
		\label{dielc}
	\end{eqnarray}

	\acknowledgments
	
	This work is supported by Mid-career Researcher Program through the National Research Foundation of Korea grant No. NRF-2021R1A2B5B02002603 and NRF-2022H1D3A3A01077468.
	DG would also like to thank Institute Postdoctoral Fellowship (IPDF) of IIT Kharagpur.

	\bibliographystyle{jhep}
	\bibliography{refpapers1.bib}

\providecommand{\href}[2]{#2}\begingroup\raggedright\begin{thebibliography}{10}

\bibitem{QGTRachel}
I.~Komissarov, T.~Holder, and R.~Queiroz, {\it The quantum geometric origin of
  capacitance in insulators},  {\em Nat. Commun.} {\bf 15} (2024) 4621.

\bibitem{QGTAhn}
J.~Ahn, G.-Y. Guo, N.~Nagaosa, and A.~Vishwanath, {\it Riemannian geometry of
  resonant optical responses},  {\em Nature Phys.} {\bf 18} (2022) 290.

\bibitem{Wang1}
Z.~Wang and S.-C. Zhang, {\it Simplified topological invariants for interacting
  insulators},  {\em Phys. Rev. B} {\bf 86} (2012) 165116.

\bibitem{RestaR}
R.~Resta, {\it The insulating state of matter: a quantum-geometric property},
  {\em Eur. Phys. J. B} {\bf 79} (2011) 121.

\bibitem{Provost:1980nc}
J.~P. Provost and G.~Vallee, {\it {RIEMANNIAN STRUCTURE ON MANIFOLDS OF QUANTUM
  STATES}},  {\em Commun. Math. Phys.} {\bf 76} (1980) 289--301.

\bibitem{10.1098/rspa.1984.0023}
M.~V. Berry, {\it Quantal phase factors accompanying adiabatic changes},  {\em
  Proceedings of the Royal Society of London. A. Mathematical and Physical
  Sciences} {\bf 392} (1984), no.~1802 45--57.

\bibitem{MaldacenaM}
J.~M. Maldacena, {\it The large n limit of superconformal field theories and
  supergravity},  {\em Int. J. Theor. Phys.} {\bf 38} (1999) 1113.

\bibitem{GKPW1}
S.~S. Gubser, I.~R. Klebanov, and A.~M. Polyakov, {\it Gauge theory correlators
  from non-critical string calculus},  {\em Phys. Lett. B} {\bf 428} (1998)
  105.

\bibitem{GKPW2}
E.~Witten, {\it Anti-de sitter space and holography},  {\em Adv. Theor. Math.
  Phys.} {\bf 2} (1998) 253.

\bibitem{HQMReV}
S.~A. Hartnoll, A.~Lucas, and S.~Sachdev, {\em Holographic Quantum Matter}.
\newblock MIT Press, 2018.

\bibitem{Iqbal:2009fd}
N.~Iqbal and H.~Liu, {\it {Real-time response in AdS/CFT with application to
  spinors}},  {\em Fortsch. Phys.} {\bf 57} (2009) 367--384,
  [\href{http://arxiv.org/abs/0903.2596}{{\tt arXiv:0903.2596}}].

\bibitem{Ghorai:2024nxs}
D.~Ghorai, T.~Yuk, Y.-K. Han, and S.-J. Sin, {\it {Classes of holographic Mott
  gaps}},  {\em JHEP} {\bf 10} (2024) 062,
  [\href{http://arxiv.org/abs/2404.10412}{{\tt arXiv:2404.10412}}].

\bibitem{Ghorai:2023vuo}
D.~Ghorai, T.~Yuk, and S.-J. Sin, {\it {Order parameter and spectral function
  in d-wave holographic superconductors}},  {\em Phys. Rev. D} {\bf 109}
  (2024), no.~6 066004, [\href{http://arxiv.org/abs/2309.01634}{{\tt
  arXiv:2309.01634}}].

\bibitem{Ghorai:2023wpu}
D.~Ghorai, T.~Yuk, and S.-J. Sin, {\it {Fermi arc in p-wave holographic
  superconductors}},  {\em JHEP} {\bf 10} (2023) 003,
  [\href{http://arxiv.org/abs/2304.14650}{{\tt arXiv:2304.14650}}].

\bibitem{Han:2024rbr}
Y.-K. Han, D.~Ghorai, T.~Yuk, and S.-J. Sin, {\it {Holographic mean field
  theory and Kondo lattice}},  {\em JHEP} {\bf 12} (2024) 123,
  [\href{http://arxiv.org/abs/2407.01978}{{\tt arXiv:2407.01978}}].

\bibitem{SangJinNature}
H.~Im et~al., {\it Observation of kondo condensation in a degenerately doped
  silicon metal},  {\em Nature Phys.} {\bf 19} (2023) 676.

\bibitem{Faulkner:2013ica}
T.~Faulkner, M.~Guica, T.~Hartman, R.~C. Myers, and M.~Van~Raamsdonk, {\it
  {Gravitation from Entanglement in Holographic CFTs}},  {\em JHEP} {\bf 03}
  (2014) 051, [\href{http://arxiv.org/abs/1312.7856}{{\tt arXiv:1312.7856}}].

\bibitem{Oh:2021xbe}
E.~Oh, T.~Yuk, and S.-J. Sin, {\it {The emergence of strange metal and
  topological liquid near quantum critical point in a solvable model}},  {\em
  JHEP} {\bf 11} (2021) 207, [\href{http://arxiv.org/abs/2103.08166}{{\tt
  arXiv:2103.08166}}].

\bibitem{Oh:2020cym}
E.~Oh, Y.~Seo, T.~Yuk, and S.-J. Sin, {\it {Ginzberg-Landau-Wilson theory for
  Flat band, Fermi-arc and surface states of strongly correlated systems}},
  \href{http://arxiv.org/abs/2007.12188}{{\tt arXiv:2007.12188}}.

\bibitem{Sukrakarn:2023ncp}
S.~Sukrakarn, T.~Yuk, and S.-J. Sin, {\it {Mean field theory for strongly
  coupled systems: Holographic approach}},  {\em JHEP} {\bf 06} (2024) 100,
  [\href{http://arxiv.org/abs/2311.01897}{{\tt arXiv:2311.01897}}].

\bibitem{Song:2019asj}
G.~Song, J.~Rong, and S.-J. Sin, {\it {Stability of topology in interacting
  Weyl semi-metal and topological dipole in holography}},  {\em JHEP} {\bf 10}
  (2019) 109, [\href{http://arxiv.org/abs/1904.09349}{{\tt arXiv:1904.09349}}].

\bibitem{HoloTopo}
M.~Byun, T.~Yuk, Y.-K. Han, D.~Ghorai, and S.-J. Sin, {\it Topology in
  holographic mean-field theory at zero and finite temperature},  {\em arXiv
  preprint} (2025) [\href{http://arxiv.org/abs/2508.01767}{{\tt
  arXiv:2508.01767}}].

\bibitem{Ge:2016lyn}
X.-H. Ge, Y.~Tian, S.-Y. Wu, S.-F. Wu, and S.-F. Wu, {\it {Linear and quadratic
  in temperature resistivity from holography}},  {\em JHEP} {\bf 11} (2016)
  128, [\href{http://arxiv.org/abs/1606.07905}{{\tt arXiv:1606.07905}}].

\end{thebibliography}\endgroup
\end{document}